\documentclass[12pt]{iopart}
\pdfoutput=1 
\usepackage{iopams,graphicx,color}
\usepackage{ifthen}
\usepackage[pdftex,breaklinks=true,bookmarks=false]{hyperref}

\usepackage[utf8]{inputenc}

\usepackage{amssymb}
\usepackage{subeqnarray}
\usepackage{ulem}
\usepackage{pgf}
\usepackage{tikz}
\usetikzlibrary{arrows}
\usetikzlibrary[shapes.geometric]

\newcommand{\journ}[5]
{\ifthenelse{\equal{#1}{pr}}{
{\it #5}, #4 Phys. Rev {\bf #2} \href{http://link.aps.org/abstract/PR/v#2/e#3}{#3}}
{\ifthenelse{\equal{#1}{prl}}{
{\it #5}, #4 Phys. Rev. Lett {\bf #2} \href{http://link.aps.org/abstract/PRL/v#2/e#3}{#3}}
{\ifthenelse{\equal{#1}{prb}}{
{\it #5}, #4 Phys. Rev. B {\bf #2} \href{http://link.aps.org/abstract/PRB/v#2/e#3}{#3}}
{\ifthenelse{\equal{#1}{prd}}{
{\it #5}, #4 Phys. Rev. D {\bf #2} \href{http://link.aps.org/abstract/PRD/v#2/e#3}{#3}}
{\ifthenelse{\equal{#1}{pra}}{
{\it #5}, #4 Phys. Rev. A {\bf #2} \href{http://link.aps.org/abstract/PRA/v#2/e#3}{#3}}
{\ifthenelse{\equal{#1}{arxiv}}{
{\it #5}, #4 \href{http://arxiv.org/abs/#2.#3}{arXiv:#2.#3}}
{\ifthenelse{\equal{#1}{rmp}}{
{\it #5}, #4 Rev. Mod. Phys {\bf #2} \href{http://link.aps.org/abstract/RMP/v#2/e#3}{#3}}
{\ifthenelse{\equal{#1}{cond-mat}}{preprint
\href{http://arxiv.org/abs/cond-mat/#2}{cond-mat/#2}}
{\ifthenelse{\equal{#1}{pre}}{
{\it #5}, #4 Phys. Rev. E {\bf #2} \href{http://link.aps.org/abstract/PRE/v#2/e#3}{#3}}}
{\it #5}, #4 #1 {\bf #2} #3}}}}}}}
}

\newcommand{\journdoi}[6]{{\it #5}, #4 #1 {\bf #2} \href{http://dx.doi.org/#6}{#3}}

\newcommand{\ra}{\rangle}
\begin{document}

\title{R\'enyi entanglement entropies in quantum dimer models : from criticality to topological order}

\author{Jean-Marie St\'ephan, Gr\'egoire Misguich and Vincent Pasquier}

\address{Institut de Physique Th\'eorique,
CEA, IPhT, CNRS, URA 2306, F-91191 Gif-sur-Yvette, France.}

\eads{\mailto{jean-marie.stephan@cea.fr}, \mailto{gregoire.misguich@cea.fr}, \mailto{vincent.pasquier@cea.fr} }

\begin{abstract}
Thanks to Pfaffian techniques, we study the R\'enyi entanglement entropies and the entanglement spectrum of large subsystems for two-dimensional
Rokhsar-Kivelson wave functions constructed from a dimer model on the triangular lattice.
By including a fugacity $t$ on some suitable bonds, one interpolates between the triangular lattice ($t=1$) and the square lattice ($t=0$).
The wave function is known to be a massive $\mathbb Z_2$ topological liquid for $t>0$ whereas it is a gapless critical state at $t=0$.
We mainly consider two geometries for the subsystem: that of a semi-infinite cylinder, and the disk-like setup proposed by Kitaev and Preskill [Phys. Rev. Lett. 96, 110404 (2006)].
In the cylinder case, the entropies contain an extensive term  -- proportional to the length of the boundary -- and a universal sub-leading constant $s_n(t)$. 
Fitting these cylinder data (up to a perimeter of $L=32$ sites) provides $s_n$ with a very high numerical accuracy ($10^{-9}$ at $t=1$ and $10^{-6}$ at $t=0.5$).
In the topological $\mathbb{Z}_2$ liquid phase we find $s_n(t>0)=-\ln 2$, independent of the fugacity $t$ and the R\'enyi parameter $n$.
At $t=0$ we recover a previously known result, $s_n(t=0)=-\frac{1}{2}\ln(n)/(n-1)$ for $n<1$ and  $s_n(t=0)=-\ln(2)/(n-1)$ for $n>1$.
In  the disk-like geometry --  designed to get rid of the boundary
contributions --  we find an entropy $s^{\rm KP}_n(t>0)=-\ln 2$ in the whole massive phase whatever $n>0$, in agreement with the result
of Flammia {\it et al.} [Phys. Rev. Lett. 103, 261601 (2009)]. Some results for the gapless  limit $R^{\rm KP}_n(t\to 0)$ are discussed.

\end{abstract}

\date{\today}

\maketitle

\tableofcontents

\section{Introduction}
\label{sec:intro}

It is now widely recognized that the entanglement entropy is a useful quantity to probe many-body quantum states.
It can be used to detect critical states in one-dimensional chains, through the celebrated logarithmic divergence \cite{hlw94,vlrk03,korepin04,cc04}.
In two dimensions it can be a used  to characterize (massive) topologically ordered  states. In particular,
it allows to distinguish a topological wave function from a more conventional disordered and featureless state.
In a gapped phase the entanglement entropy of a large subsystem contains a contribution proportional to the length (in two dimensions) of its boundary plus a subleading
term $S_{\rm topo}$ which contains some information about the nature of the phase. In a state with topological order, this subleading term is related to the total quantum dimension, that is to the
content in elementary excitation \cite{hiz05,kp06,lw06}.
This idea has been successfully applied to some fractional quantum hall states \cite{haque07,fl08,lbsh10}.
Extracting  the subleading term in lattice models is not a trivial task \cite{kp06,lw06}
but it was first shown to be feasible using quantum dimer wave functions on the triangular lattice \cite{fm07}.
Since the work of Moessner and Sondhi \cite{ms01} this type of states have been intensively studied since they
offer some rather simple realization of topologically ordered states with non trivial finite-size effects and finite correlation length
(contrary to toric-code like models \cite{kitaev03,msp02}).

In this paper we also consider some dimer wave functions -- named after Rokhsar and Kivelson (RK) \cite{rk88} -- which are linear superposition of fully packed dimer coverings on the triangular lattice.
By including a fugacity
on some suitable bonds, one continuously interpolates between the triangular lattice ($t=1$) and the square lattice ($t=0$).  
In the triangular case the wave function is known to be a massive $\mathbb Z_2$ topological liquid \cite{ms01,fms02,iif02} whereas it
 is a gapless critical state at $t=0$ \cite{rk88}.
Exploiting previous results \cite{fm07,stephan09} on the reduced density matrix (RDM) of RK states, 
we can obtain not only the entanglement entropy but also the full entanglement spectrum on large systems.
Using extensively the Pfaffian formulation of the classical dimer partition function \cite{kasteleyn61},
 as well as some perturbation theory for determinant \cite{fisher63,fms02} we perform calculations
{\it in the thermodynamic limit} while keeping the boundary length finite.

In the cylinder geometry we can treat the  infinite height limit and perimeters up to $L=32$ ($38$ at $t=0$).
In the disk-like geometry proposed by Kitaev and Preskill \cite{kp06}, we perform exact calculation for disks of
 radii up to $\rho \simeq 4.5$ lattice spacings embedded in a {\it infinite} system, therefore extending significantly the previous
 entanglement calculations on triangular dimer wave functions \cite{fm07}. 
This technique allows to confirm the value $S_{\rm topo}=-\ln(2)$ with  high precision in the whole
 massive phase (not only at the triangular point $t=1$). This value turns out not to depend on the R\'enyi parameter, in agreement with
the argument by Flammia {\it et al.} \cite{fhhw09}.
We also discuss the structure of the {\it entanglement spectrum}, showing that it contains a non-degenerate ``ground-state''
and a gap. In Sec.~\ref{sec:espectrum}, a micro-canonical point of view is used to relate the density of states of the entanglement spectrum and the  R\'enyi entanglement entropies.

When $t=0$ the dimers
are restricted to the bonds of a square lattice.
Although non-generic,\footnote{They correspond to fine tuned multi-critical points \cite{vbs04,aff04,plf07}.}  such critical RK wave-functions
associated to some conformally invariant critical points
are useful since they offer one of the very few situations
where one can study the entanglement in a critical wave-functions in more than one dimension \cite{fm06,hsu09,stephan09,oshikawa10,hsu10b}. 
Another point of view is that, for long cylinder geometries, the entanglement in these two-dimensional systems is related to the Shannon entropies in -- now generic --  quantum critical chains \cite{stephan09,smp10,zbm11,smp11}.
The sub-leading constant in the cylinder geometry depends
on the compactification radius \cite{hsu09,stephan09,oshikawa10,hsu10b} and shows a singularity at some critical value of the R\'enyi parameter \cite{smp11}.
The result in a Kitaev-Preskill geometry is less clear and we discuss our numerical results at the end of Sec. \ref{sec:kp}.

\section{Entanglement entropy as a Shannon entropy}

After a brief introduction to 
dimer RK wave-functions \cite{rk88}, we
review how one can construct the RDM and Schmidt decomposition for these states.

\subsection{Rokhsar-Kivelson wave functions}\label{ssec:RK}
\begin{figure}[!ht]
 \begin{center}
\includegraphics{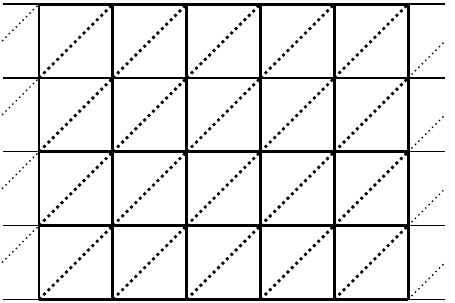}
 \end{center}
\caption{Triangular lattice with cylindrical boundary conditions ($L_x=6$, $L_y=5$).
 Each ``diagonal link'' (dotted lines) has fugacity $t$, the others have fugacity 1.
}
\label{fig:triangular_lattice}
\end{figure}

We start from a classical two-dimensional hard-core dimer model on a triangular lattice, with fugacity $t$ on 
``diagonal'' links (Fig.~\ref{fig:triangular_lattice}). This fugacity allows to interpolate between the square lattice ($t=0$)
 and the isotropic triangular lattice ($t=1$).

 The classical partition function of this system reads 
\begin{equation}
 \mathcal{Z}=\sum_c e^{-E(c)}=\sum_c \displaystyle{t^{ \textrm{\# diagonal dimers}}},
\end{equation}
where the sum runs over all dimer coverings $c$. When  $t=0$ (square lattice),
 the model is known to be critical \cite{fisher63,kasteleyn63}, its long range behavior is described by a compact free field \cite{fhros04,alet05}. Otherwise it has a finite correlation length \cite{ms01,fms02,iif02}.
An Hilbert space is then constructed by associating a basis state $|c\rangle$ to each  classical dimer configuration $c$. 
Different classical configurations correspond to orthogonal states.
The RK wave function is the normalized linear combination
 of all basis states with an amplitude equal to the square root of the classical weight~:
\begin{equation}\label{eq:RK}
 |RK\rangle=\frac{1}{\sqrt{\mathcal{Z}}}\sum_c e^{-E(c)/2} |c\rangle.
\end{equation}

Following Henley \cite{henley04} one can construct some local Hamiltonians for which Eq.~\ref{eq:RK} is an exact ground state, but the
precise form of these Hamiltonians will not be used in the following.

\subsection{R\'enyi entanglement entropy}
We divide the system into two parts $A$ and $B$.
Each subsystem is a set of bonds, and its degrees of freedom are the corresponding dimer occupancies.
The RDM
of $A$ is obtained by tracing over the degrees of freedom in $B$:
\begin{equation}
 \rho_A=\textrm{Tr}_B |RK\rangle \langle RK|
\end{equation}
Then, the R\'enyi entanglement entropy is defined as
\begin{equation}
S_n=\frac{1}{1-n}\ln \textrm{Tr } \rho_A^{\,n},
\end{equation}
 where $n$ is not necessarily an integer.
 Two limits are of interest. For $n \to 1$, $S_n$ reduces to the Von Neumann entanglement entropy :
\begin{equation}
 S_1=S^{\textrm{vN}}=-\textrm{Tr }\rho_A \ln \rho_A
\end{equation}
For $n \to \infty$, only the largest  eigenvalue $p_{\rm max}$ of the RDM matters :
\begin{eqnarray}
 S_{\infty}&=&-\ln p_{\rm max}.
\end{eqnarray}
This quantity is also called single copy entanglement.
 To compute all the R\'enyi entropies, we need all the eigenvalues of the RDM. 
In the following, we shall see that calculating each eigenvalue amounts to solving a combinatorial problem. The
 procedure has been discussed in details elsewhere \cite{fm07,stephan09} and is recalled below for completeness.
\subsection{Schmidt decomposition}
\label{sec:sd}

\begin{figure}[!ht]
\includegraphics{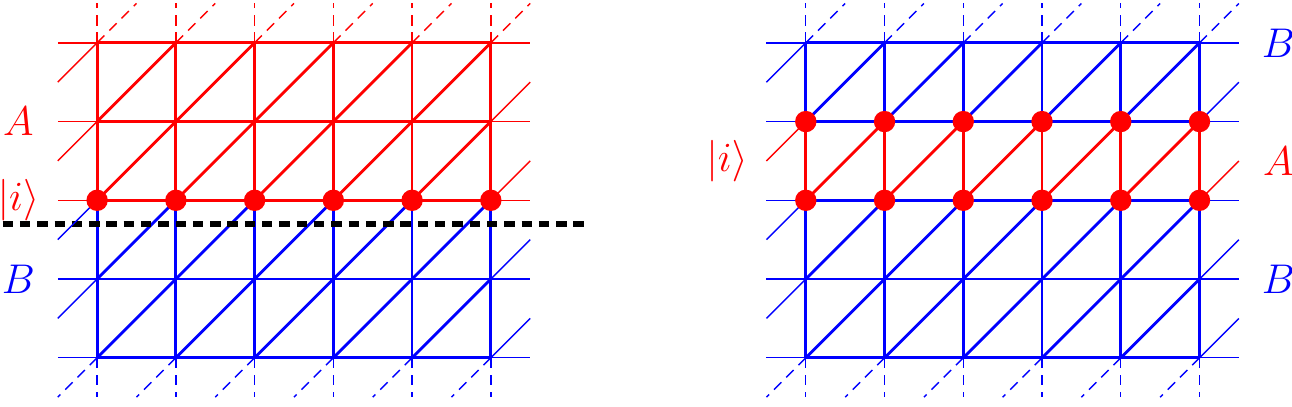}
\caption{(Color online) Partition of the lattice
in two subsystems A (red bonds) and B (blue bonds).
Left: the subsystems A and B are semi-infinite cylinders.
Boundary sites are marked by filled red circles.
Each boundary site can either be occupied by a dimer in $A$ (spin $\uparrow$), or a dimer in $B$ (spin $\downarrow$).}
\label{fig:cylinder_cut}
\end{figure}

We consider the geometry of an infinite cylinder cut into two parts, as in the left of Fig.~\ref{fig:cylinder_cut}.
 The reasoning is the same for the other geometries we considered.
 The sites which touch a bond in A and an bond in B (red circles in Fig.~\ref{fig:cylinder_cut}) are called boundary sites.
 We assign a spin $\sigma_j$ to each boundary site : $\sigma_j=\uparrow$ if the site is occupied by a dimer in $A$, $\sigma_j=\downarrow$ if it is occupied by a dimer in $B$.
 We denote by 
\begin{equation}
|i\rangle=|\sigma_1,\sigma_2,\ldots,\sigma_{L_x}\rangle
\end{equation}
the whole spin configuration at the boundary.

Now, let $\mathcal{E}_i^A$ (resp. $\mathcal{E}_i^B$) be the set of dimer configurations in $A$ (resp. $B$) compatible with $|i\rangle$ at the boundary.
 Thanks to the hardcore constraint, they share no common element :
\begin{equation}
 \mathcal{E}_i^A \cap \mathcal{E}_{i^\prime}^{B}=\emptyset\qquad,i\neq i'
\label{eq:ortho}
\end{equation}
 Each configuration $c$ can be written as
\begin{equation}
 c=a\cup b\quad,a\in \mathcal{E}_i^A\quad,b\in \mathcal{E}_i^B
\end{equation}
and the energy decomposed as
\begin{equation}
 E(c)=E_A(a)+E_B(b)
\end{equation}
This allows to write the RK state as :
\begin{equation}\label{eq:RKbis}
 |RK\rangle=\frac{1}{\sqrt{\mathcal{Z}}}\sum_i \left[\sum_{a\in \mathcal{E}_i^A}e^{-E_A(a)/2}|a\rangle\right]
\times\left[\sum_{b\in \mathcal{E}_i^B}e^{-E_B(b)/2}|b\rangle\right]
\end{equation}
Defining a new normalized set of RK states in $A$ and $B$
\begin{eqnarray}
 |{\rm RK}_i^A \ra &=& \frac1{\sqrt{\mathcal{Z}_i^A}} 
 \sum_{a\in {\cal E}_i^A} e^{-\frac{1}{2} E_A(a)}|a\ra, 
    \\
 |{\rm RK}_i^B \ra &=& \frac1{\sqrt{\mathcal{Z}_i^B}}
 \sum_{b\in {\cal E}_i^B} e^{-\frac{1}{2} E_{B}(b)}|b\ra,
   \\
 \textrm{with} &~&  
 \mathcal{Z}_i^{\Omega}=\sum_{\omega\in \mathcal{E}_i^\Omega}
e^{-E_\Omega(\omega)}
 ~~(\Omega=A,B),
\end{eqnarray}
Eq.~(\ref{eq:RK}) becomes
\begin{equation}
 |RK\rangle=\sum_i \sqrt{p_i}|RK_i^A\rangle |RK_i^B\rangle,
\label{eq:schmidt}
\end{equation}
with 
\begin{equation}\label{eq:proba}
 p_i=\frac{\mathcal{Z}_i^A \mathcal{Z}_i^B}{\mathcal{Z}}.
\end{equation}
Eq.~\ref{eq:schmidt} is actually the Schmidt decomposition of the RK state (the orthogonality of the Schmidt vectors is guarantied by Eq.~\ref{eq:ortho}),
and the $\{p_i\}$ are the eigenvalues of the RDM:  
\begin{equation}
 \rho_A=\sum_i p_i |RK_i^A\rangle \langle RK_i^A|,
\end{equation}
This way, one can obtain the R\'enyi entropy :
\begin{equation}
 S_n=\frac{1}{1-n}\ln \left(\sum_i p_i^{\,n}\right) \label{eq:Sn}.
\end{equation}
The entanglement entropy calculation has been reduced to finding some probabilities in the classical dimer problem. In the next section we will show that, using standard Pfaffian techniques, one can
 obtain exact formulae for the $p_i$. 

\section{Classical probabilities}
\subsection{Pfaffian}
The Pfaffian of a ($2n\times 2n$) antisymmetric matrix $M$ is defined as 
\begin{equation}\label{eq:pfaff}
 \textrm{Pf }M=\sum_{\pi \in S_{2n}}' \epsilon(\pi)M_{\displaystyle{\pi_1 \pi_2}}M_{\displaystyle{\pi_3\pi_4}}\ldots M_{\displaystyle{\pi_{2n-1}\pi_{2n}}},
\end{equation}
where $\epsilon(\pi)$ denotes the signature of a permutation $\pi$. The sum runs over all permutations
 of $\{1,2,\ldots,2n\}$ satisfying the constraints 
\begin{equation}
 \begin{array}{ccccl}
  \pi_{2i-1}<\pi_{2i}&&,&&1<i<n\\
\pi_{2i-1}<\pi_{2i+1}&&,&&1<i<n-1
 \end{array}
\end{equation}
A very important relation is 
\begin{equation}
 \left(\textrm{Pf } \,M\right)^2=\det M,
\end{equation} 
It is especially useful because it allows to compute the Pfaffian numerically in a time proportional to $n^3$ using standard determinant routines (and sometimes analytically).
\subsection{Kasteleyn theory}

The problem of enumerating dimer configurations on a planar lattice is a classic combinatorial problem, which was solved
independently by Kasteleyn \cite{kasteleyn61} and Temperley and Fisher \cite{fisher61}. We consider the case $t=1$ for simplicity
but the generalization to any $t$ is straightforward. For any planar graph, the partition function (number of dimer coverings) is given by

\begin{equation}\label{eq:Kasteleyn_theorem}
\mathcal{Z}=\left|\textrm{Pf } \mathcal{K}\right|,
\end{equation}
where $\mathcal{K}$ is an antisymmetric matrix constructed as follows.  Putting arrows on all the links, a matrix element of $\mathcal{K}$ is
\begin{equation}\label{eq:adjacency}
\mathcal{K}_{ij}=\left\{
\begin{array}{cll}
+1&& \textrm{ if the arrow points from $i$ to $j$}\\
-1&& \textrm{ if the arrow points from $j$ to $i$}\\
0&& \textrm{   if $i$ and $j$ are not nearest neighbors}
\end{array}
\right.
\end{equation} 

The Kasteleyn matrix must also satisfy the {\it clockwise-odd rule} : the product
of the arrows orientations ($\pm 1$) around any elementary plaquette (running clockwise) has to be $-1$.
Kasteleyn showed that i) such a matrix $\mathcal{K}$ exists for any planar graph and ii) it insures that all  terms in the sum have the same sign (the signature of the permutation always compensate
that of the product of matrix elements). It is immediate to check that ii) implies Eq~\ref{eq:Kasteleyn_theorem}.

\begin{figure}[!ht]
\begin{center}
\includegraphics{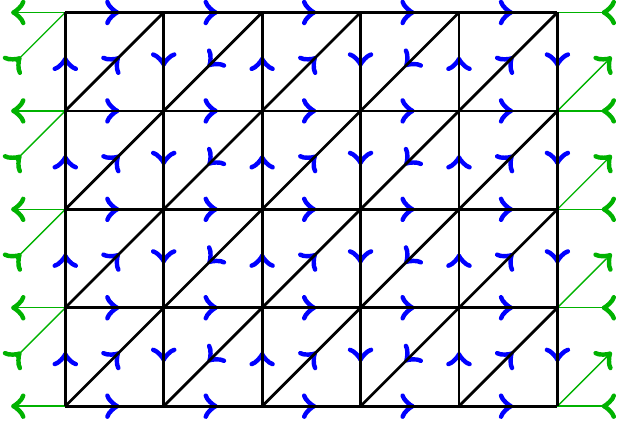}
\caption{Kasteleyn orientation of the $(L_x=6, L_y=5)$ lattice (a weight $t$ is given to ``diagonal'' links). Blue arrows: orientation of the bonds. 
Green : bonds present because of periodic boundary conditions along the $x-$ axis (see Ref.~\cite{mccoywu}). Their orientations are reversed compared to their
``bulk'' counterparts.}
\label{fig:cyl_orientation}
\end{center}
\end{figure}

A Kasteleyn matrix obeying Eq.~\ref{eq:Kasteleyn_theorem} can also be found for cylindrical boundary conditions.
An example for the triangular lattice with cylindrical boundary conditions\footnote{In the case of toroidal boundary condition the situation is slightly more complicated, and the number of dimer covering is given by a linear combination
of four Pfaffians, see Ref.~\cite{mccoywu} for more details.} is shown
in Fig.~\ref{fig:cyl_orientation}.

In the following we will demonstrate how each probability $p_i$ can be computed as
 a determinant, taking the example of the cylinder geometry. 
\subsection{Classical probabilities}
\begin{figure}[!ht]
 \begin{center}
\includegraphics{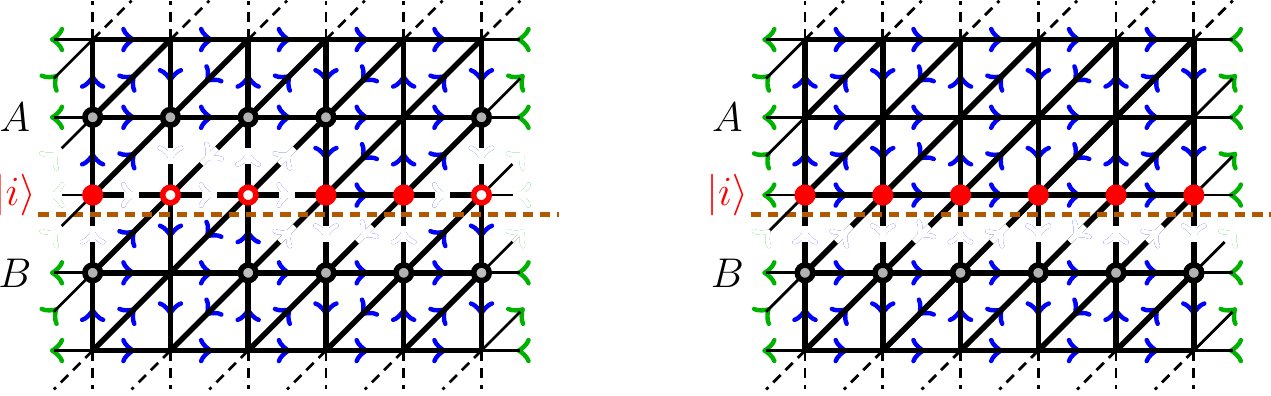}
 \end{center}
\caption{Two examples with $|i\rangle=|\!\!\uparrow \downarrow \downarrow \uparrow \uparrow \downarrow\rangle$
 on the left, and $|i\rangle=|\!\!\uparrow \uparrow \uparrow \uparrow \uparrow \uparrow\rangle$ on the right.
 Filled red circle:  boundary site occupied by a dimer in $A$ (spin $\uparrow$).
 Empty red circle: boundary site occupied by a dimer in $B$ (spin $\downarrow$).
 To ensure that a boundary site be occupied by a dimer in $A$ (resp. $B$),
 all edges in $B$ (resp. $A$) coming from this site have to be removed. Notice that after the removal,
 $A$ and $B$ are disconnected. black circles filled in grey are sites which are connected to a boundary site through
 a link that has been removed. As explained in the text, the size of the determinant is given by the number of circles.
 $p_i$ is therefore a $16 \times 16$ determinant for the configuration on the left and a $12 \times 12$ determinant for
 the configuration on the right. 
}
\label{fig:Kast_kill}
\end{figure}

To find the probabilities of Eq.~\ref{eq:proba}, we need
 to compute $\mathcal{Z}_i^A \mathcal{Z}_i^B$, which is the partition function restricted to dimer configurations compatible
 with the boundary spin configuration $|i\ra=|\sigma_1,\ldots \sigma_{L_x}\rangle$. It can be evaluated as the Pfaffian of a modified Kasteleyn matrix
\begin{equation}
 \mathcal{Z}_i^A \mathcal{Z}_i^B =\textrm{Pf } \mathcal{K}^{(i)}
\end{equation}
where $\mathcal{K}^{(i)}$ is deduced from $\mathcal{K}$ by removing the appropriate links in a simple way.
 If $\sigma_j=\uparrow$, a dimer emanating from the boundary site $j$ has to be in $A$, and we remove links in $B$ emanating from site $j$.
 If $\sigma_j=\downarrow$ we remove links in $A$ emanating from site $j$. See Fig.~\ref{fig:Kast_kill} for two examples, one with the boundary configuration
 $|i\rangle=|\!\uparrow\downarrow\downarrow\uparrow\uparrow\downarrow\ra$ and one with $|i\rangle=|\!\!\uparrow\uparrow\uparrow\uparrow\uparrow\uparrow\rangle$. The computation of any such probability apparently requires the ratio of two
 $L_xL_y\times L_xL_y$ determinants. However, using a known trick \cite{fisher63}, the computation can be greatly simplified.

\subsection{Perturbation theory for determinants in an infinite system} 
Following Ref.~\cite{fisher63}, $p_i^2$ may be written as
\begin{equation}
p_i^2=\det (1+\mathcal{K}^{-1}\mathcal{E}^{(i)})\quad,\quad \mathcal{E}^{(i)}=\mathcal{K}^{(i)}-\mathcal{K}
\end{equation}
The important point is that the matrix element $\mathcal{E}^{(i)}_{\mathbf{r}\mathbf{r'}}$ is non zero only
 if the link $\mathbf{r} \leftrightarrow \mathbf{r'}$ has been removed.
 Then, a matrix element of $\mathcal{K}^{-1}\mathcal{E}^{(i)}$ is
\begin{equation}
 \left(\mathcal{K}^{-1}\mathcal{E}^{(i)}\right)_{\mathbf{r}\mathbf{r'}}=\sum_{\mathbf{s}}
\mathcal{K}^{-1}_{\mathbf{r}\mathbf{s}}\mathcal{E}^{(i)}_{\mathbf{s}\mathbf{r'}}.
\end{equation}
It is non-zero only if $\mathbf{r'}$ is a site belonging to a removed link. We name these sites ``vicinity sites'', and
 they of course depend on the boundary configuration $|i\rangle$. A boundary site is automatically a vicinity site, but the converse is not true however.
If we denote by $E_i$ the set of vicinity sites and by $n_i$
 their number, $\mathcal{K}^{-1}\mathcal{E}^{(i)}$ is a $L_x L_y\times L_x L_y$ matrix, but only
 $n_i$ columns are non identically zero. Then, using the antisymmetry of the determinant, any cell with
 indices $\mathbf{r}$ and $\mathbf{r'}$ not both in $E_i$ can be set to zero
 by appropriate linear combinations of rows and columns.
 Therefore, the determinant may be computed as its restriction to the sites in $E_i$. 
\begin{equation}\label{eq:perturbation_trick}
 p_i^2=\det \left(\left(1+\mathcal{K}^{-1}\mathcal{E}^{(i)}\right)_{|_{E_i}}\right)
\end{equation}
This so called ``perturbation theory for determinants'' has been previously used in Ref.~\cite{fisher63}
 to compute exactly the monomer-monomer correlation on
 the square lattice in the thermodynamic limit ($L,L_y \to \infty$), and further extended in Ref.~\cite{fms02} to the triangular lattice.
 For computational purpose this is a huge simplification, because the size of the determinant has been reduced from $L_xL_y$ to $n_i \sim \mathcal{O}(L_x)$,
 and the total system we are interested in can be infinite ($L_y\to\infty$). Contrary also to the transfer matrix approach \cite{stephan09},
 this method allows us to treat any shape of boundary. This will be particularly useful while studying the geometry proposed by Kitaev and Preskill \cite{kp06}. 

 For this to work we also need to compute exactly certain matrix elements of
 the inverse Kasteleyn matrix $\mathcal{K}^{-1}$. This can be done  using standard Fourier and integral techniques, see \ref{sec:green}. 

Let us now specify the case of the (infinitely long) cylinder geometry cut into two parts. An example of spin configuration is
 shown in Fig.~\ref{fig:Kast_kill}, where boundary sites are represented by red circles (filled or empty depending on the spin).
 Other vicinity sites are circles filled in grey. It is easy to check that $2L_x\leq n_i \leq 3 L_x$ for all boundary configurations.
 Since there are a priori $2^{L_x}$ boundary configurations and each probability is of complexity $\sim n_i^3$, the R\'enyi entropy can be evaluated
 in a time $\sim L_x^3\times 2^{L_x}$. This allow us to go to relatively large system sizes of order $L_x\sim 30$.

\section{Results for the infinite cylinder}\label{sec:cyl}

When the height $L_y$ is infinite, the entropies $S_n$ only depend on the perimeter $L_x=L$. As usual, the leading term is
non universal and scales with $L$, and we are interested in the first subleading contribution $s_n$:
\begin{equation}
 S_n(L)\simeq \alpha_n L + s_n + o(1)
\end{equation}

\subsection{Topological entanglement entropy and R\'enyi index}
\label{sec:topo_renyi}
\begin{figure}
 \centering
\includegraphics[width=12cm]{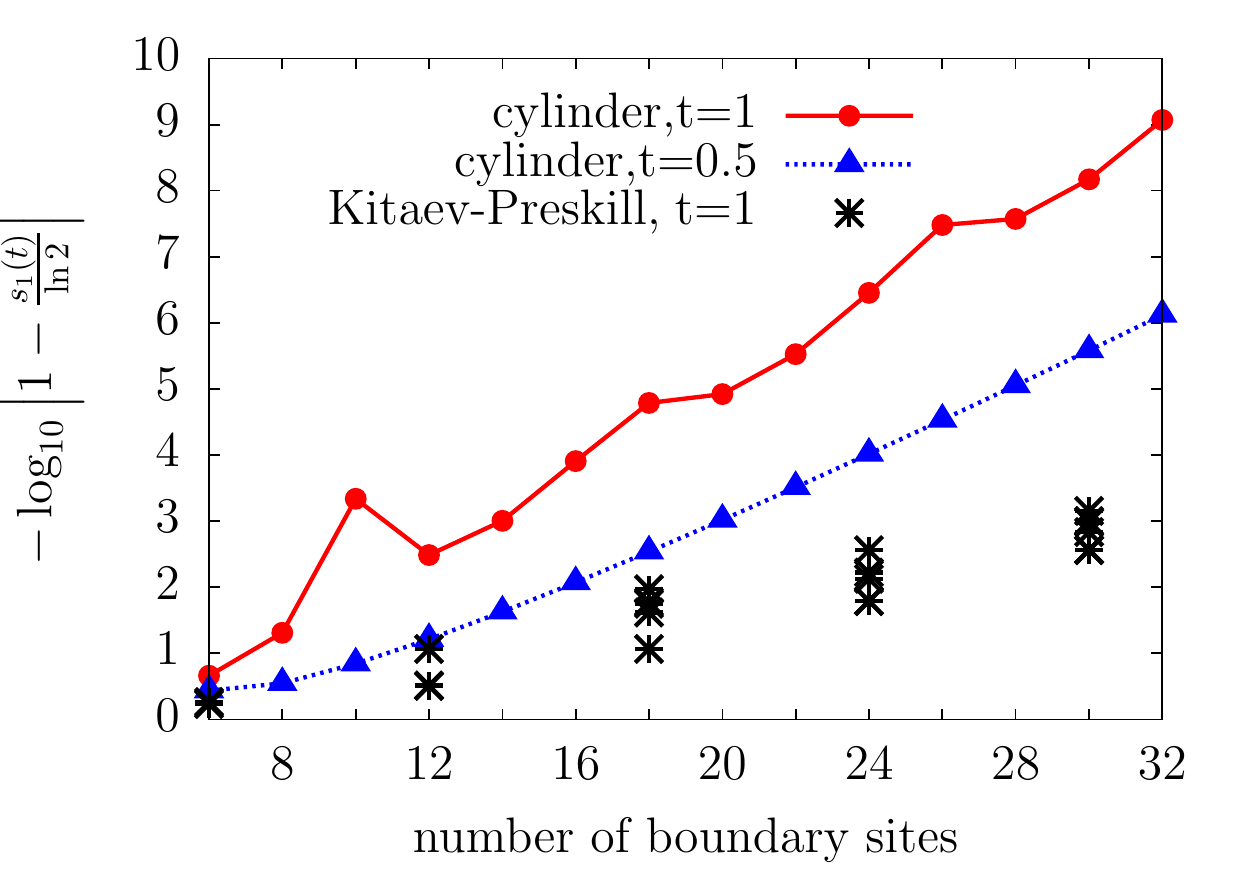}
\caption{Number of correct digits in the numerical estimate of the topological constant, as a function of the number of boundary sites. 
For the cylinder geometry 
we show the data for $t=1$ (red circles) and $t=0.5$ (blue triangles). The number of boundary sites is just $L$ in this case, and the estimate
 is obtained by a fit to $aL+s_1$ for two even consecutive values of $L$. 
 The convergence to the correct value is exponentially
 fast, with an effective correlation length close to the dimer-dimer correlation length (which can have an imaginary part\cite{fms02,iif02}, hence the oscillations we observe). For comparison we also show the data in the Kitaev-Preskill geometry, slightly anticipating on Sec.~\ref{sec:kp}.
 }
\label{fig:convergence}
\end{figure}

For gapped topological wave functions, the subleading  constant $s_1$ in the Von Neumann entropy
has been shown to be related to the content of the phase in terms of fractionalized particles, and to the total quantum dimension $D$ in particular \cite{kp06,lw06}:
$s_1=-\ln(D)$.
In the original works the subleading constant $s_1$ was extracted by combining the entropies of different subsystems in a planar geometry. We show here that the subleading term can 
be extracted in a -- somewhat simpler -- cylinder geometry  (see also \cite{lbsh10}).

For $t>0$ the present dimer wave-functions realize 
the simplest topological phase, the so-called $\mathbb{Z}_2$ liquid with quantum dimension $D=2$.
One therefore expects to have $s_1=-\ln 2$ in the whole topological phase. So far, this has only been checked numerically at $t=1$ \cite{fm07}.
In addition, Ref.~\cite{fhhw09} argues that this topological entanglement entropy is independent of the R\'enyi index  $n$. 
We present here some results for infinitely high cylinders for various values of  $t$ and $n$, which support this result.
 The convergence to the topological entropy is exponentially fast, as can be seen in Fig.~\ref{fig:convergence}. For
 generic values of $t$ and $n$, this allows us to
 get this constant with a very high accuracy: for example at $t=1$ our best estimate is $|s_1(t=1)+\ln 2|\simeq 10^{-9}$.
 It is widely believed that in massive phases the topological entropies (subleading terms) are independent of short-range correlations, but this is not {\it proven}. The present results, which strongly indicate that $s_n=-ln(2)$ for any $t>0$, therefore brings some additional support to the robustness of topological entropies.
 In general finite-size effects get larger when increasing $n$ at fixed $t$, and it is more advisable to numerically study
 low-$n$ R\'enyi entropies. However, as is shown in \ref{sec:pmax}, the calculation for $n\to \infty$ simplifies greatly,
 and the result $s_{\infty}(t>0)=-\ln 2$ can even be obtained rigorously. We further discuss this result in Sec.~\ref{sec:renyi_lbf}. 
\begin{figure}[!ht]
\begin{center}
 \includegraphics[width=12cm]{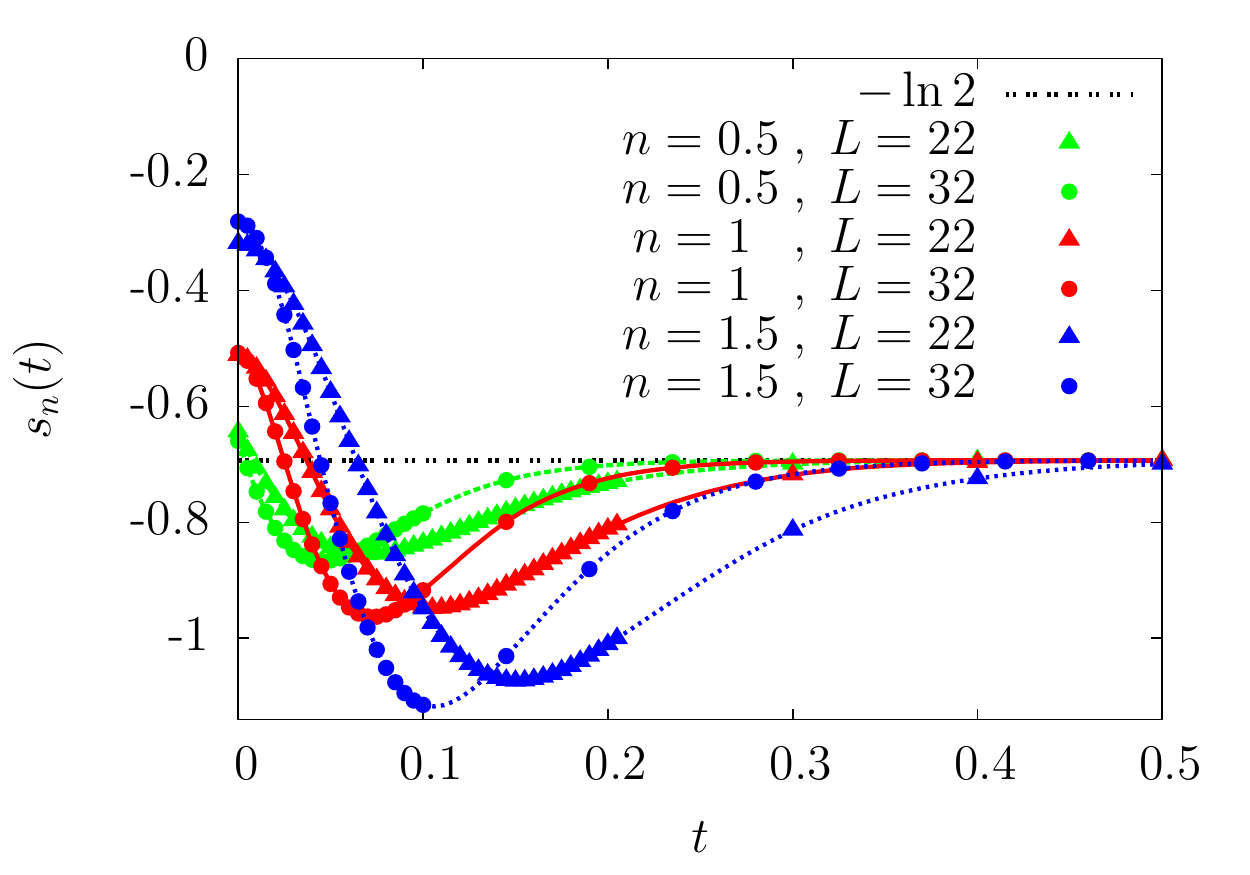}
\end{center}
\caption{Sub-leading constants $s_n(t)$ for $3$ different values of the R\'enyi parameter ($n=0.5,1,1.5$).
For each $t$ and $n$,  $s_n(t)$ is extracted from $S_n(L)$ using two consecutive even values of $L$ (up to $L=32$).
In the thermodynamic limit the results are expected to converge to $s_n(t)=-\ln 2$ for all $n>0$ and $t>0$.}
\label{fig:snt}
\end{figure}
At fixed $n$ the convergence is also less clear when $t$ is small since the correlation length $\xi(t)$ diverges when approaching $t=0$ and the finite-size effects become very important when $L\gtrsim \xi(t)$. Still,
the curve $s_n(t>0)$ approaches $-\ln 2$ when  $L\to\infty$. The data plotted in Fig.~\ref{fig:snt} are indeed compatible with $s_n(t)=-\ln 2$ for all $n=0.5,1,2$ and $t>0$.
  The scaling close to $t=0$ will be discussed later in Sec.~\ref{ssec:t=0}.

\subsection{Thermodynamical entropy}

\begin{figure}[!ht]
\begin{center}
 \includegraphics[width=12cm]{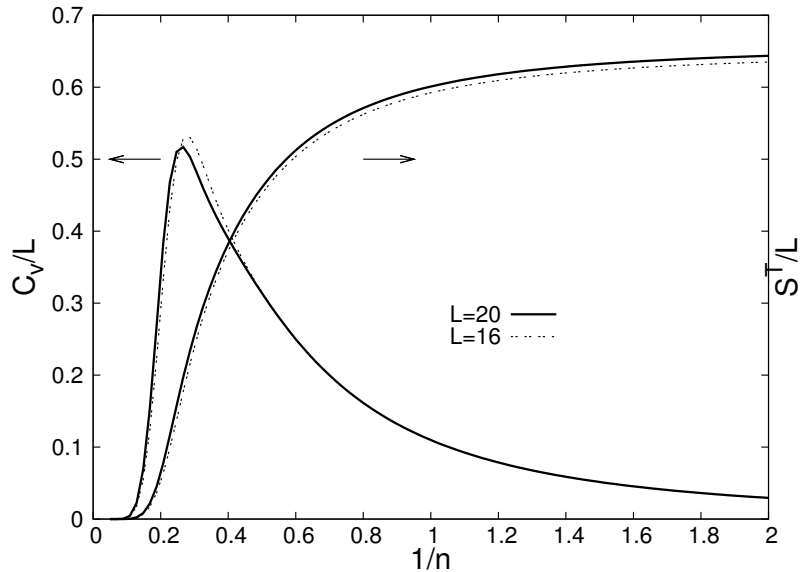}
\end{center}
\caption{Thermodynamic entropy per site $S^T_n/L$  (monotonously increasing, right axis) and its associated ``specific heat'' (peaked at $n\simeq0.25$, left axis) $C_v=-n dS^T/dn$.
Fugacity $t=1$.}
\label{fig:s_cv}
\end{figure}

\begin{figure}[!ht]
\begin{center}
 \includegraphics[width=12cm]{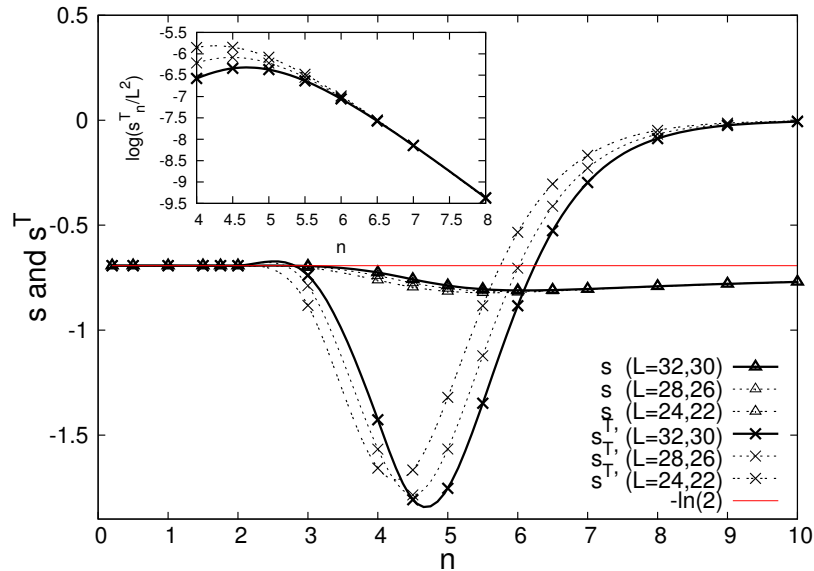}
\end{center}
\caption{Large $n$ behavior of the subleading constant $s_n(t=1)$ of the R\'enyi entropy, and $s^T_n(t=1)$, the subleading constant of the thermodynamical entropy.
They both give $-\ln(2)$ for small $n$, but differ for large $n$. This is a finite-size-effect: as shown in the inset,
$s^T\sim L^2 \exp(-\Delta n)$ for large $n$. We thus have $s\simeq s^T=\simeq -\ln(2)$ up to  $n\sim \ln(L)$.
}
\label{fig:s_st}
\end{figure}

The behavior for large values of the R\'enyi index $n$ is displayed in Fig.~\ref{fig:s_st} (triangular dots).
Although it is roughly constant and close to $-\ln(2)$, due to the finite-size of the system there are some visible deviations for $n\gtrsim 3$.
This is even more visible if we consider a slightly different entropy, $S^T_n$, defined as:
\begin{eqnarray}
 S^T_n&=&\left(1-\partial_n\right) \ln\left(Z_n\right) \\
Z_n&=&\sum_i p_i^n
\end{eqnarray}
which can also be written as the Shannon entropy associated to the normalized probabilities $\tilde p_i$:
\begin{equation}
 S^T_n=-\sum_i \tilde p_i \ln(\tilde p_i) \;\;{\rm with}\;\; \tilde p_i=\frac{p_i^n}{Z_n}.
\end{equation}
Both entropies match at $n=1$ ($S^T_{n=1}=S_{n=1}$)
and are simply related otherwise: $S^T_n=(1-n\partial_n)((1-n)S_n)$.
The ``thermodynamic'' entropy $S^T$ has also a leading term $\mathcal{O}(L)$ and a sub-leading term, $s^T_n$.
The extensive (and non-universal) part is plotted in Fig.~\ref{fig:s_cv} as a function of the ``temperature'' $1/n$.
To stress the similarity with usual statistical mechanics, we also plotted the associated ``specific heat'' defined as  a derivative of $S^T$: $C_v=-n\frac{dS^T}{dn}$.

The sub-leading term $s^T_n$ is plotted in Fig.~\ref{fig:s_st} (crosses). It is very close to $-\ln(2)$ at small $n$, but goes
to $s^T=0$ when $n\to\infty$. This is indeed expected since the thermodynamic entropy $S^T_{n=\infty}$ -- which corresponds to zero ``temperature'' -- is equal to the log of the degeneracy of the configuration with the highest probability, which is non-degenerate in our case.
However, the crossover from $-\ln(2)$ to $0$ takes place at values of $n$ which are larger and larger when $L\to\infty$. This
can be checked in the inset of Fig.~\ref{fig:s_st}, where the numerical data appear to be correctly fitted by
\begin{eqnarray}
 s^T_{n \gg \ln(L)} &\sim& L^2 \exp(-n \Delta) \\
s^T_{n \ll \ln(L)} &\sim& -\ln(2) 
\end{eqnarray}
where $\Delta\simeq 1.32$ is the entanglement gap at $t=1$. We finally note that the  calculation
of $p_{\rm max}$ given in Sec.~\ref{sec:pmax} proves rigorously that 
$\lim_{L\to\infty}\lim_{n\to\infty} s_n=-\ln(2)$ and $\lim_{L\to\infty}\lim_{n\to\infty} s^T_n=0$.

\subsection{Scaling when $t\to0$ and $L\to\infty$ with fixed $L\cdot t$}
\label{ssec:t=0}

The critical point $t=0$ has already been studied \cite{stephan09,smp11} and is known to give:
\begin{equation}
 s_n(0)=\left\{
\begin{array}{ccc}
 \ln R-\frac{\ln n}{2(n-1)}&,&0< n\leq 1\\
\frac{n}{n-1}\ln R&,& n>1
\end{array}
\right.
\label{eq:sn0}
\end{equation}
\begin{equation}
 s^T_n(0)=\left\{
\begin{array}{ccc}
 \ln \left(\sqrt{n}R\right)-\frac{1}{2}&,&0< n\leq 1\\
0,& n>1
\end{array}
\right.,
\label{eq:sTt0}
\end{equation}
where the compactification radius is  $R=1$ (free fermions) for the present dimer wave-functions, but could be tuned by adding some dimer-dimer interactions \cite{alet05}.

\begin{figure}[!ht]
\begin{center}
 \includegraphics[width=12cm]{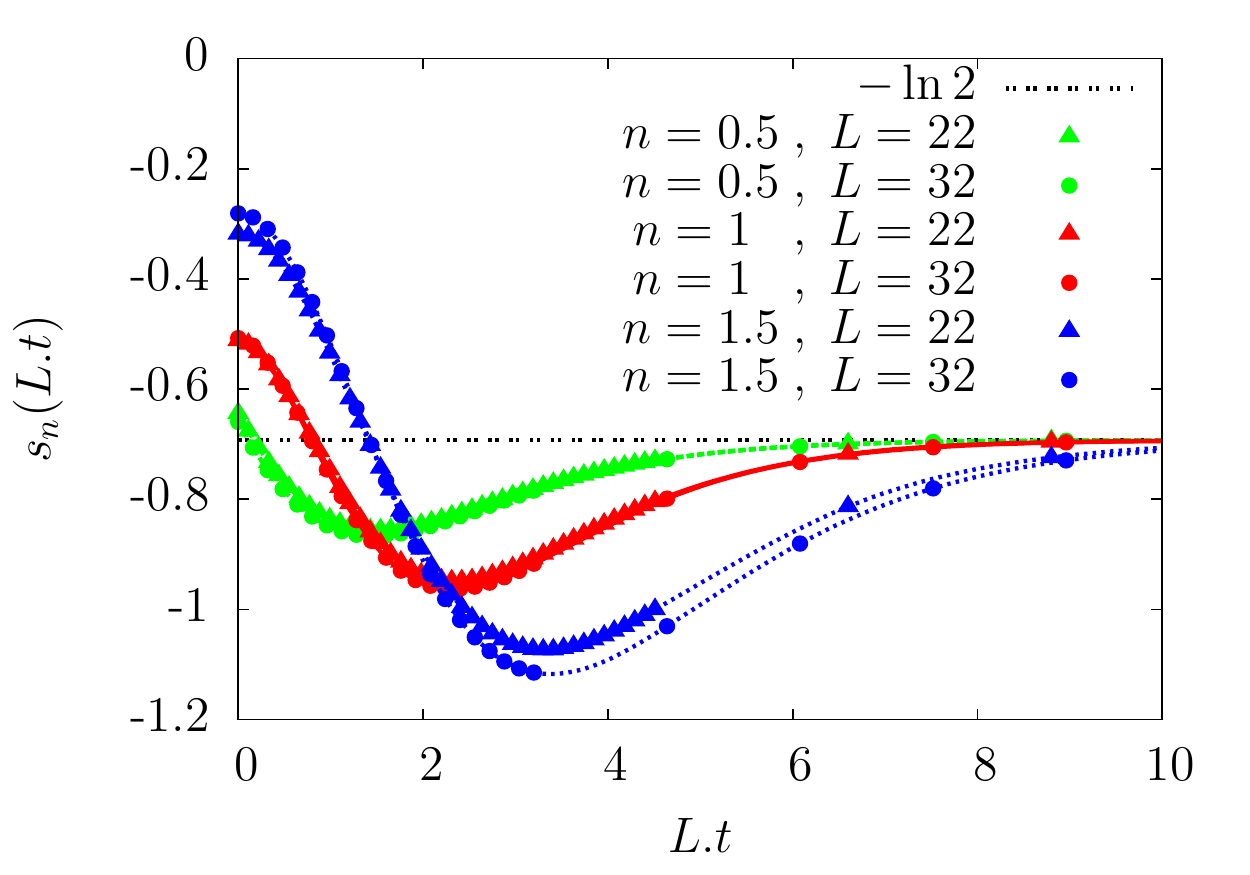}
\end{center}
\caption{Sub-leading constants $s_n$, as a function of  $t\times L$. For each values of $n$, the data
corresponding to different values of $t$ and $L$ appear to be well described by a function of $t\times L$ only.}\label{fig:cyl_scaling}
\end{figure}

The correlation length $\xi(t)$  diverges
as $\xi(t)\sim t^{-1}$ when $t\ll 1$ \cite{fms02}. In Fig.~\ref{fig:cyl_scaling} we plot the subleading constant $s_n(t,L)$ as
a function of $L\cdot t\simeq L/\xi(t)$. It appears that, for a given value of $n$, the data curves corresponding to different
values of $t$ and $L$ approximately collapse onto each other. This shows that, when
the system size $L$ is much bigger than the correlation length $\xi(t)\sim t^{-1}$,
we find the correct topological entanglement entropy $s_n= -\ln(2)$. On the other hand, when $L$ is of the same order of magnitude
than $\xi(t)$ (and much larger than the lattice spacing)  $s_n$ turns out to be some non-trivial function of $n$ and  $L \cdot t$.
When $L \cdot t \to 0$ the system effectively behaves as a critical system of dimers on a square lattice and $s_n$ converges to 
Eq.~\ref{eq:sn0}, as expected.

\subsection{Entropy of a zig-zag line}

\begin{figure}[!ht]
\begin{center}
 \includegraphics[width=12cm]{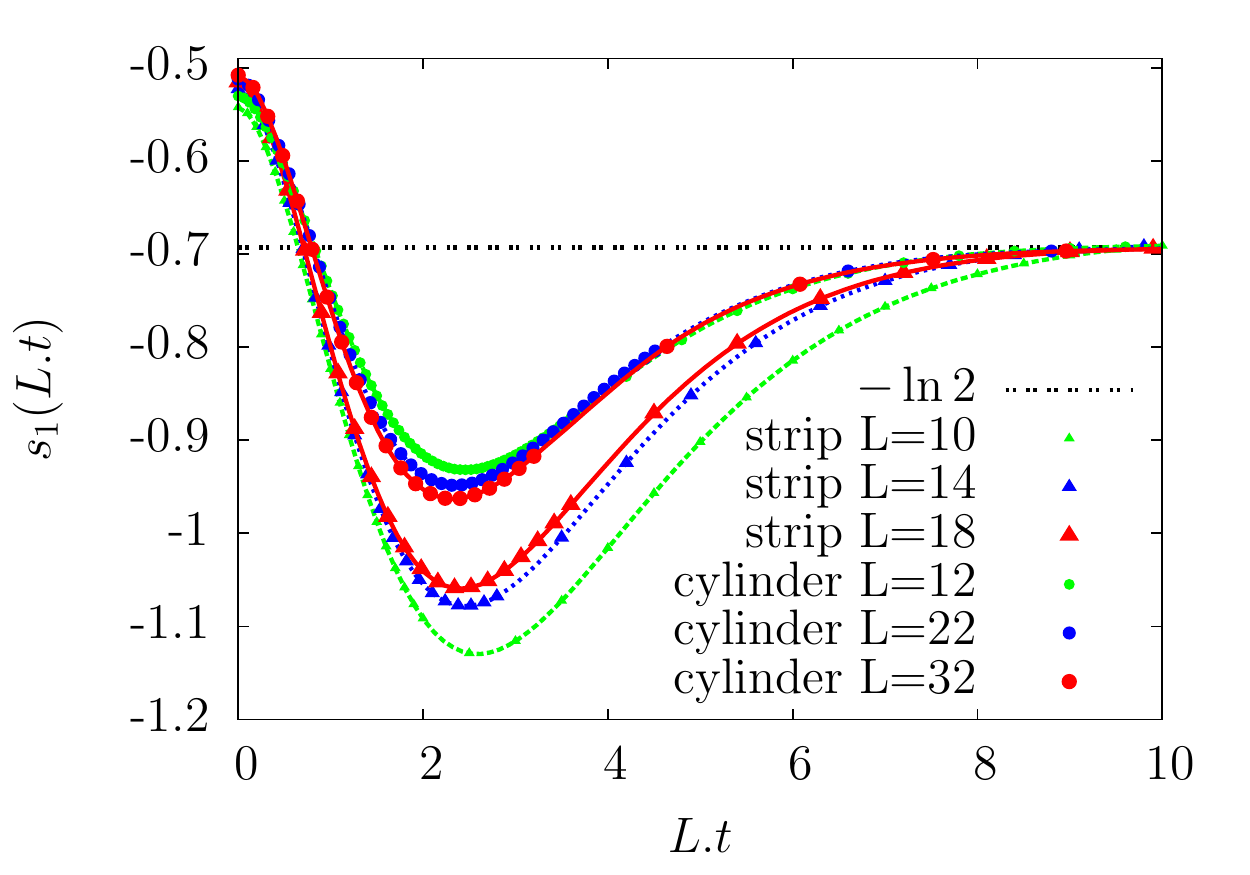}
\end{center}
\caption{Sub-leading constants for the entanglement entropy calculated numerically in two geometries : half-infinite cylinder and zig-zag strip (see text).}
\label{fig:cyl_vs_strip}
\end{figure}

As explained in Sec.~\ref{sec:sd}, the eigenvalues of the RDM
of a half infinite cylinder are the classical probabilities of the ``spin'' configurations
$|i\ra = |\sigma_1,\sigma_2,\ldots,\sigma_{L}\rangle$. But one may also consider a  zig-zag
line and the probabilities $p_\alpha$ of the dimer configurations on that lines.
The ``spins'' are now replaced by the dimer occupancies (say 0 or 1) of the zig-zag bonds.
Theses probabilities can be computed using exactly the same perturbed-Pfaffian method as before.
However, in terms of entanglement, the entropy we compute is that of a the ''zig-zag'' chain
shown in the right of Fig.~\ref{fig:cyl_vs_strip}. Although the probabilities are computed in a very similar way, this calculation does not describe
the entanglement of a two-dimensional subsystem, but that of a one-dimensional line winding around the cylinder.

The associated entropies, already considered in Ref.\cite{fm07}, have a leading term proportional to $L$ and a subleading
contribution of order $\mathcal{O}(L^0)$. The results, plotted in Fig.~\ref{fig:cyl_vs_strip}, show
that the subleading constant $s_1$ has a dependence on $t$ and system size $L$ which is very similar to that of the half-cylinder entropy.
It is possible that, as a function of $L\cdot t$,  the zig-zag line and half-infinite cylinder  converge to the same curves for sufficiently large $L$. In any case, the zig-zag results
clearly converges to $-\ln(2)$ in the thermodynamic limit for $t>0$.

One may ask if the zig-zag entropy would also give access to the quantum dimension for a {\it general} topologically ordered wave-function (not of RK type, and even not based on dimers).
We believe that it is {\it not} the case. The present dimer RK states enjoy a special property: once the dimer occupancies are fixed along the zig-zag chain, the upper and lower half-cylinders are completely decoupled. For this reason, the entropy of the zig-zag chain is very close to that of a half cylinder. This would not hold for more generic states and a {\it thick} strip (sufficiently large compared to the correlation length) would be probably required to access the quantum dimension in general.

\subsection{Infinite R\'enyi and bipartite fidelity}
\label{sec:renyi_lbf}
As already emphasized, the infinite-$n$ R\'enyi limit selects the largest eigenvalue of the RDM, which is the probability
 of the most likely configuration in the dimer language:
\begin{equation}
 S_{\infty}=-\ln p_{\rm max}
\end{equation}
 For the cylinder geometry the corresponding boundary configuration 
 $|i_{\rm max}\rangle$ is particularly simple (see Fig.~\ref{fig:Kast_kill} for a graphical representation):
\begin{equation}
 |i_{\rm max}\rangle=|\!\uparrow \uparrow \ldots \uparrow\,\rangle,
\end{equation}
 and $p_{\rm max}$ can be expressed as a ratio of simple partition functions:
\begin{equation}
p_{\rm max}=\lim_{L_y \to \infty}\frac{\left[Z_{\rm cyl}(L_x,L_y/2)\right]^2}{Z_{\rm cyl}(L_x,L_y)},
\end{equation}
 where $Z_{\rm cyl}(L,h)$ is the partition function for dimers on a finite cylinder of length $L$ and height $h$.
As detailed in \ref{sec:renyi_infinity}, we then find the following  expression for $S_\infty$
\begin{equation}
 S_{\infty}=
-\sum_{k=\frac{(2m-1)\pi}{L}}^{1\leq m \leq L/2}
\ln \left(  \frac{1}{2} +\frac{1}{2}\frac{\sin^2 k-t \cos k}{\sqrt{t^2+\sin^2 k+\sin^4 k}}\right),
\end{equation}
from which one can extract the sub-leading constant
\begin{equation}
 s_{\infty}(t)=\left\{
\begin{array}{ccc}
 0\quad &,&t=0\\
-\ln 2\quad &,& t>0.
\end{array}
\right.
\end{equation}
This result has already been mentioned in Sec.~\ref{sec:topo_renyi}. 
The entropy $ S_{\infty}$ can also be considered from a different point of view.
$|RK\rangle$ is the ground state of the Rokhsar-Kivelson Hamiltonian,
 and lives on a cylinder of length $L$ and height $h$. This Hamiltonian may be written as
\begin{equation}\label{eq:lbf_hamiltonian}
 H=H_{A\cup B}=H_A+H_B+H_{A,B}^{\rm (int)},
\end{equation}
where $H_A$ (resp. $H_B$) is the Rokhsar-Kivelson Hamiltonian restricted to sites in $A$ (resp. $B$). We have
 $[H_A,H_B]=0$ and $H_{A,B}^{\rm (int)}$ contains all the interactions between $A$ and $B$.
 If we denote by $|A\rangle$ (resp. $|B\rangle$) the ground-state of $H_A$ (resp. $H_B$), $|A\otimes B\rangle=|A\rangle \otimes |B\rangle$
 the ground-state of $H_A+H_B$ and by
 $|A\cup B\rangle=|RK\rangle$ the ground state of $H_{A\cup B}$, then $p_{\rm max}$ can be reformulated as
\begin{equation}\label{eq:bf_pmax}
 p_{\rm max}=\left|  \langle A\cup B|A\otimes B\rangle\right|^2
\end{equation}
Taking minus the logarithm we get
\begin{equation}\label{eq:lbf_renyi}
 S_{\infty}=-\ln \left|\langle A\cup B|A\otimes B\rangle\right|^2
\end{equation}
The r.h.s of Eq.~\ref{eq:lbf_renyi} has been studied in Ref.~\cite{lbf} under the name logarithmic bipartite fidelity(LBF).
The (infinite) R\'enyi entanglement entropy and the LBF are  a priori not related, but we find that they are
simply equal for this particular RK wave-function. In other words, performing
 a Schmidt decomposition on the total wave function $|A\cup B\rangle$, the Schmidt state 
with the highest Schmidt value is nothing but the ground state of $H_A+H_B$, the RK Hamiltonian
where all interactions between $A$ and $B$ were switched off.

However, this relation does  not  hold exactly in general.
For instance, in the Kitaev-Preskill or Levin-Wen geometry the boundaries are not straight and in that case the boundary dimer configuration $|i_{\rm max}\ra$
is not as simple as for the cylinder. Still, as pointed out in \cite{lbf}, the equivalence between the LBF and $S_\infty$ can hold for some more complex topological states such as the
 string nets states constructed by Levin and Wen \cite{lw06}. We expect that for a generic ({\it i.e.} non RK) gapped
state, the sub-leading term in the LBF and $S_\infty$ should be the same in the thermodynamic limit (although, due to some mismatch at short distances, the extensive terms will differ).
The argument is as follows: starting from a string net state where the correspondence works, we adiabatically modify the wave function toward the state we are interested in (without closing the gap).
Doing so it is natural to expect that only the short-distance properties of the entanglement will be modified (hence the $\sim L$ term)  but not the sub-leading constant $s_\infty$ which is expected to be free from the contribution of local correlations. Although the robustness to changes in local correlations is is not
proven in general, we provide in \ref{sec:renyi_infinity} a rigorous proof that the
subleading term $s_\infty$
is equal to $-\ln 2$ in the whole massive phase of the model ($t>0$).

\subsection{Entanglement gap and entanglement spectrum}
\label{sec:espectrum}
\begin{figure}[!ht]
\begin{center}
  \includegraphics[width=8cm]{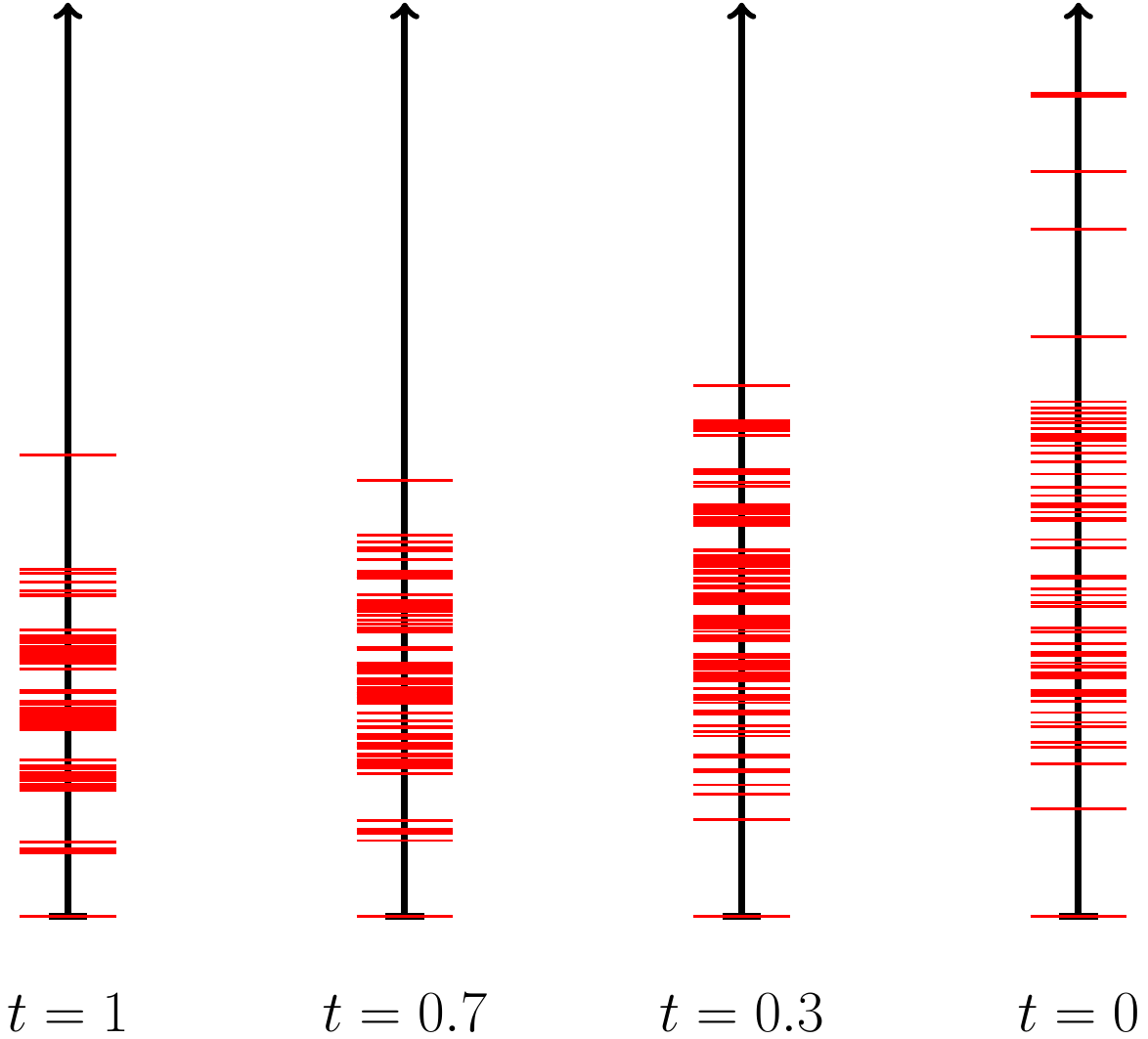}
\end{center}
\caption{Entanglement spectrum for $L=12$ and for $t=1,0.7,0.3,0$ from left to right.}
\label{fig:spect}
\end{figure}

The {\it spectrum} of the RDM contains some rich information
about the system. Looking at such spectra has been particularly fruitful in the context of the quantum Hall effect (QHE), where the entanglement spectrum
was shown \cite{lh08} to reflect some properties of the chiral gapless excitations which can propagate along an edge \cite{wen90}.
With RK wave functions the RDM eigenvalues are simple classical probabilities and we
thus have a relatively easy access to the entanglement spectra of large systems. 
\begin{figure}[!ht]
\begin{center}
  \includegraphics[width=12cm]{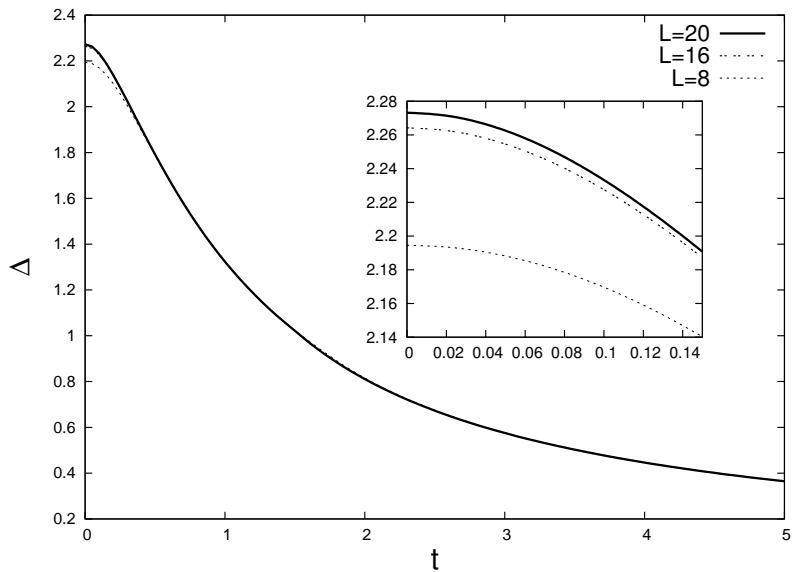}
\end{center}
\caption{Entanglement gap as a function of $t$. It is maximum at $t=0$ (square lattice)
a decreases slowly to zero when $t\to \infty$. Except very close to $t=0$ (inset) the curves for $L=16$ and 20 are practically indistinguishable at the scale of the figure, signaling negligible finite-size-effects.}
\label{fig:gap}
\end{figure}

Such spectra shown in Figs.~\ref{fig:spect}-\ref{fig:gap}, where the probabilities  $p_i$ have been converted to ``energies'': $E_i=-\ln(p_i/p_{\rm max})$.
The  first observation is that these spectra have a unique ground-state and a gap $\Delta=E_1$ to the first ``excitation''. This is true not only
in the $\mathbb Z_2$ liquid ($t>0$) but also for the critical RK wave function at $t=0$.
So, contrary to the QHE where a well defined set of low energy levels are separated from the rest \cite{lh08,tsrb10}, there is no apparent low-energy structure in the spectrum but a single ``ground state''.
One could have naively expected the entanglement gap to close when reaching the critical point at $t=0$, but this is not the case. As can be seen in Fig.~\ref{fig:gap}, the entanglement gap remains finite all the way from $t=0$ to $t=1$ (it vanishes only at $t=\infty$)
We have for instance $\Delta=1.32314$ at $t=1$ (exponentially fast convergence as a function of $L$) and $\Delta=2\ln(\pi)\simeq 2.29$ at $t=0$.\footnote{This analytical result
for $\Delta$ in the thermodynamic limit of the square lattice can be obtained by noticing that the configuration with the highest probability is $|\!\uparrow\uparrow\cdots\uparrow\rangle$ while
the next configuration has two consecutive flipped spins $|\!\uparrow\uparrow\downarrow\downarrow\uparrow\cdots\rangle$.
One can check that, for $t=0$, the ratio $p_1/p_{\rm max}$ of these two probabilities is nothing but the square of the probability for a bond located at the edge of a semi infinite square lattice to be occupied by a dimer. The latter probability has been computed in Ref.~\cite{fisher63} and is equal to $1/\pi$, which gives
$\Delta=-\ln(p_1/p_{\rm max})=2\ln(\pi)$.}
A possible interpretation is the following: the entanglement spectrum is indeed related to the spectrum of the excitations that would propagate along an edge. However, in the dimer systems we consider, there are no gapless edge excitations, even though the bulk may be gapless for $t=0$.

In the thermodynamic limit, it is possible to adopt a microcanonical point of view where the entropy $S(e)$ is simply related to the density of states:
\begin{equation}
 S(e)=\ln(\rho(e))
\end{equation}
with
\begin{equation}
 \rho(e)=\sum_i \delta(e-E_i/L).
\end{equation}
Knowing the entropy $S(e)$ from the spectrum, the energy $e(n)$ can be obtained
as a function of the R\'enyi index $n$ by inverting 
\begin{equation}
\frac{dS}{de}=n(e).
\end{equation}
The entropy $S_{n}$ is then obtained as
\begin{equation}
S_{n}=\ln(\rho(e(n))).
\end{equation}
We conclude that, for sufficiently large $L$ the entropy only depends of the density of states at some high energy $E=L\cdot e(n)$ in the spectrum.

\begin{figure}[!ht]
\begin{center}
  \includegraphics[width=7.6cm]{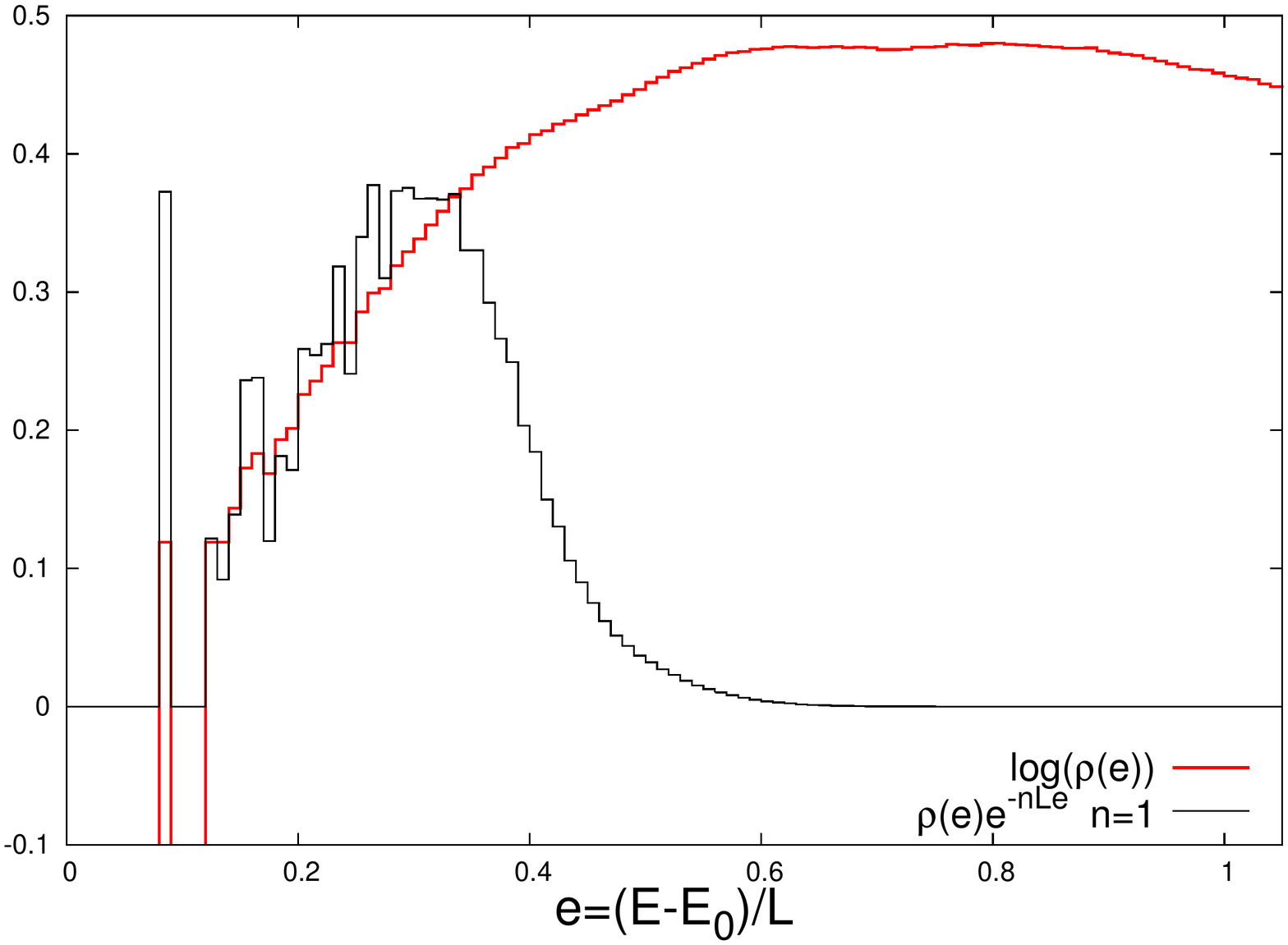}
\includegraphics[width=7.6cm]{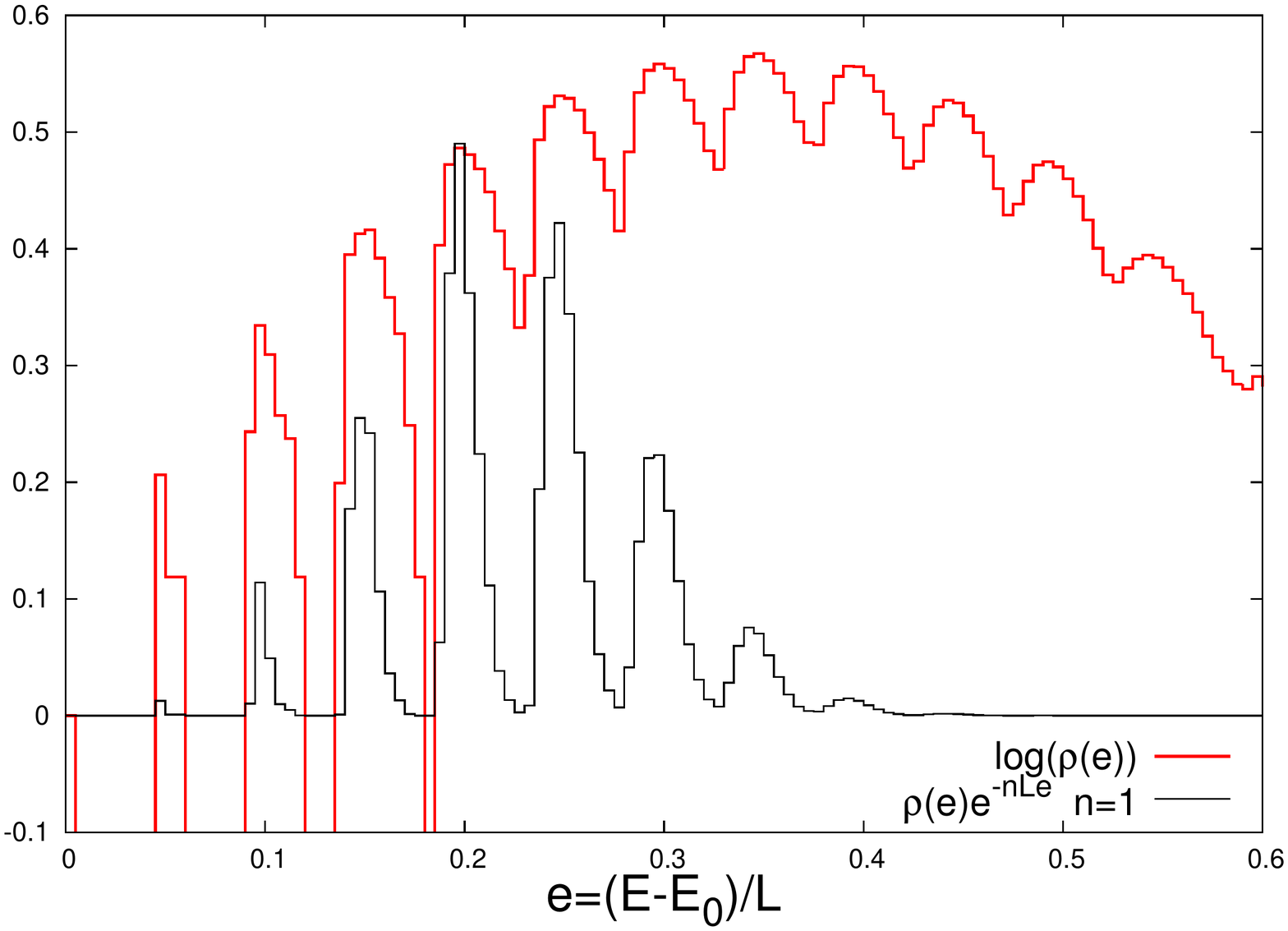}
\end{center}
\caption{Logarithm of the density of states $\rho$ associated to the entanglement spectrum of a half-infinite cylinder, as a function of the ``energy`` per site $e=(E-E_0)/L$ (arbitrary units). Top: $t=0$ (square lattice), bottom: $t=1$ (triangular lattice). To display the energy range which contribute to the Von Neumann entropy $S_1$, the probability  distribution $p(e)\sim \rho(e)\exp(-n e L)$ is also plotted for $n=1$.
System size: $L=28$. 
}
\label{fig:dos}
\end{figure}

The microcanonical entropy per site $S(e)/L$ is displayed in Fig.~\ref{fig:dos} for the triangular and square lattice (half-infinite cylinders with $L=28$). Some (finite-size)
oscillations are visible in the triangular case, and can be interpreted as the successive energy ``bands'' corresponding to $0,2,4,\cdots$ spin flips in the boundary state.
These oscillations will be smeared out in larger systems however.

\section{Long strip geometry}

\begin{figure}[!ht]
\begin{center}
  \includegraphics[width=7cm]{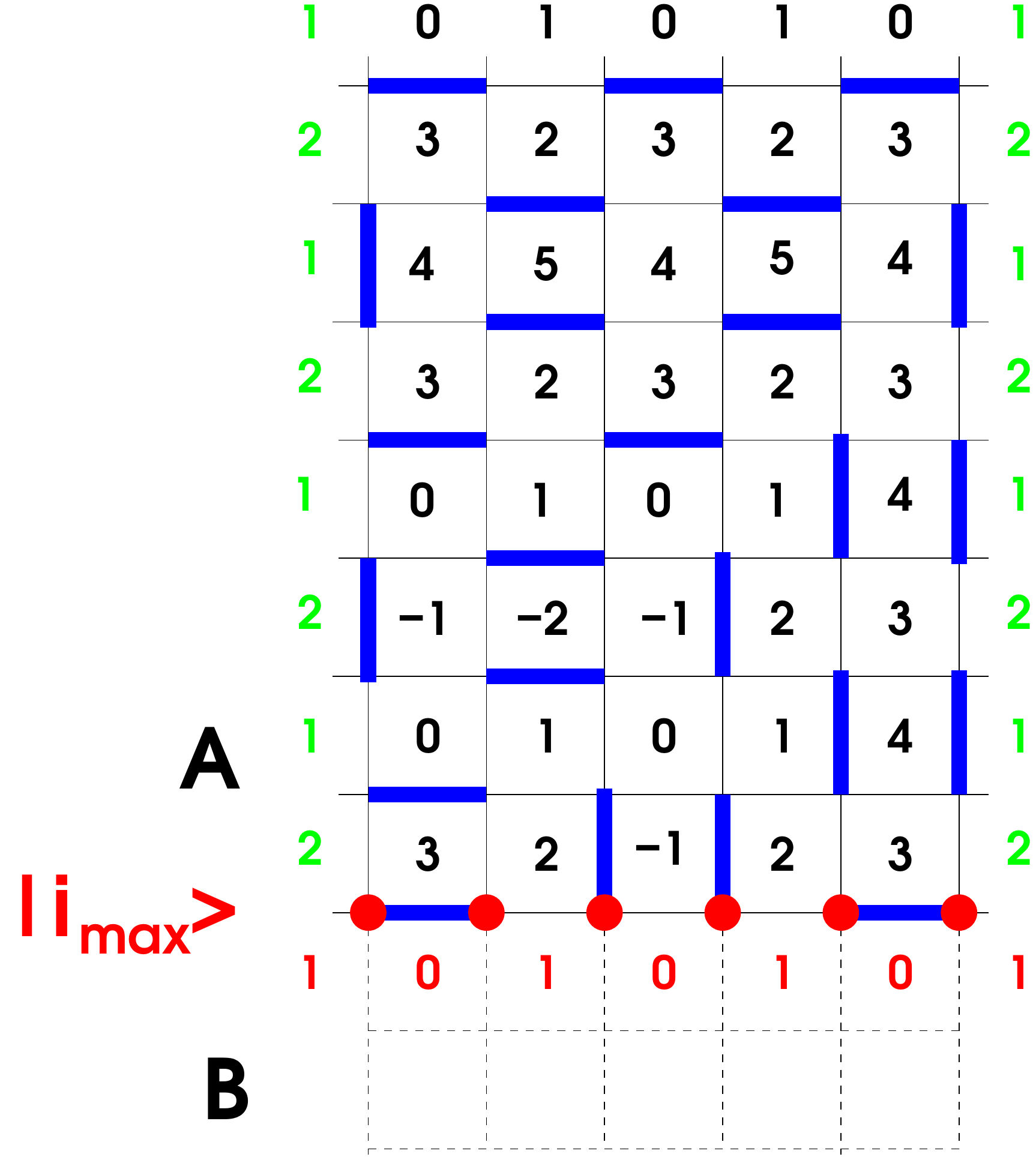}
\end{center}
\caption{Configuration $|i_{\rm max}\rangle$ with the maximal probability on the square lattice and a compatible  dimer covering of the rectangular region $A$.
The microscopic heights are indicated in units  of  $\frac{1}{2} \pi r$. When turning clockwise around a site of the even (resp. odd) sublattice the height changes by +1 (resp. -1) when crossing an empty bond, and changes by -3 (resp +3) when crossing a dimer.
The lower horizontal boundary of $A$ has a coarse-grained height which is ``flat'', with an average height equal to $\frac{1}{2}(0+1)=\frac{1}{2}$ (red). The vertical boundaries have a coarse-grained height equal to $\frac{1}{2}(1+2)=\frac{3}{2}$ (green). In the continuum limit there is an height shift $\delta=\pm \frac{1}{2}\pi r$ at each corner of $A$.
}
\label{fig:shift}
\end{figure}

The triangular lattice can also be constructed  with open boundary conditions in the $x$ direction. The geometry is no longer that of a cylinder but a long strip.
In such a situation the leading term in the entropy is still proportional to the width of the strip $L_x=L$, but the sharp corners also contribute to the sub-leading
 constant and it is not possible to extract the topological entropy for $t>0$. The critical case is more interesting, because the first subleading
 correction is now a logarithm of the width.
 The later was originally predicted to be $-\ln(L)/4$ by Fradkin and More \cite{fm06} (an application  of the Cardy-Peschel formula\cite{cp88} which describes the universal logarithmic contribution of sharp corners to the free energy in a CFT). These terms have recently been observed numerically in the closely related Shannon entropy of open critical spin chains \cite{zbm11,smp11}.
\begin{figure}[!ht]
 \centering
\includegraphics[width=12cm]{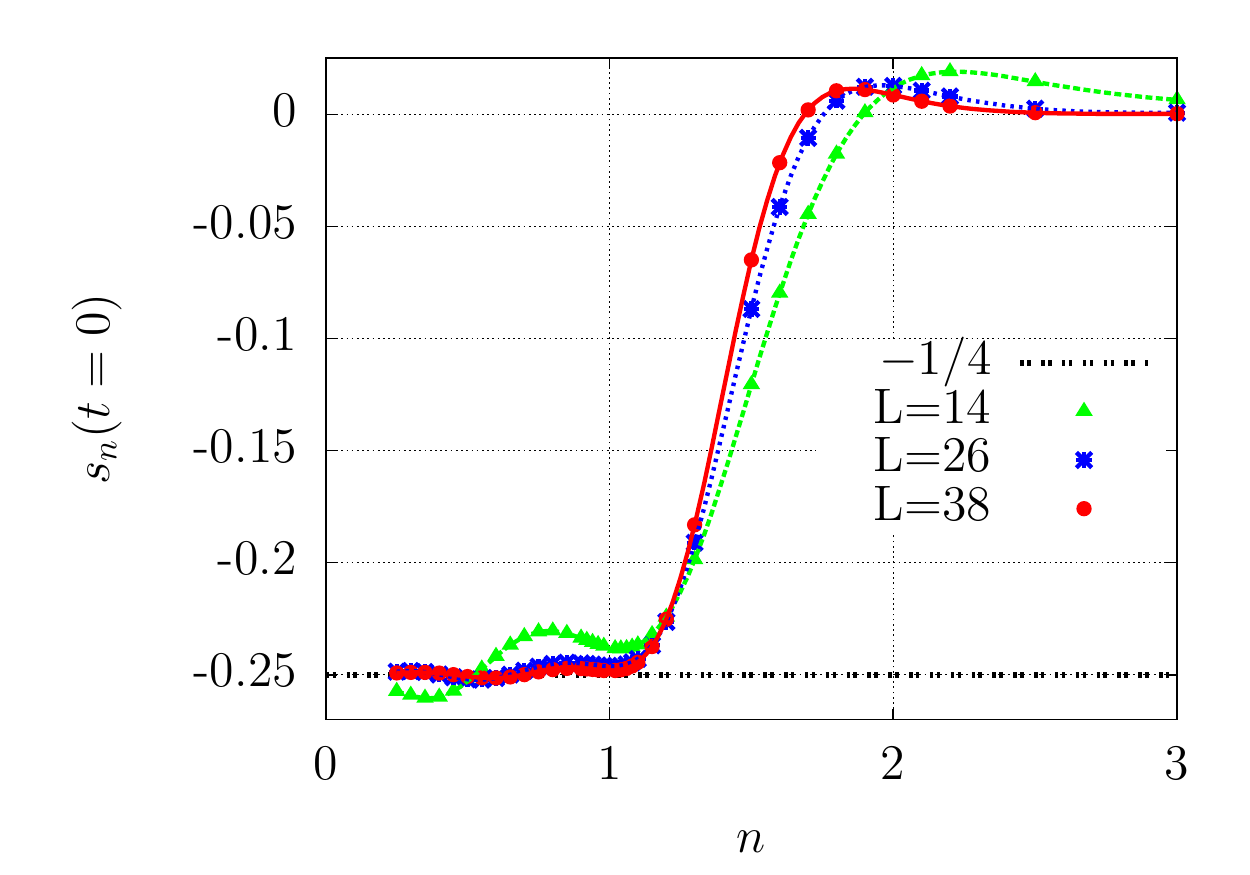}
\caption{Coefficient of the logarithmic term in the R\'enyi entropy for the strip geometry, as a function of the R\'enyi parameter $n$. This term is extracted
 from a fit $S_n=aL+b\ln L+c+d/L$ on the systems sizes $L-6,L-4,L-2,L$. Three values $L=14,26,38$ are shown.
 The data is consistent with the CFT results. For $n\leq n_c=1$, the logarithmic contribution is approximately $\sim -0.25$ (see \cite{stephan09}). For $n>n_c$
 it is close to zero as discussed in Ref.~\cite{smp11}.} 
\label{fig:renyi_band}
\end{figure}
In Fig.~\ref{fig:renyi_band} we show the coefficient of the $\ln(L)$ term
as a function of the R\'enyi index $n$ for the square lattice dimer wave function with open boundary conditions. The prediction of
Fradkin and More,  $-\frac{1}{4}$, is verified up to $n\simeq 1$. For larger values of $n$ the logarithmic term vanishes. This is a manifestation of the boundary phase transition discussed in
Ref.~\cite{smp11}. Indeed, above $n_c$ the compactness of the height field can no longer be ignored since a 
vertex operator $\cos(d h /r)$ (with $d$ an integer) becomes relevant at the boundary.
The value of $d$ can be obtained by looking at the microscopic configuration $|i_{\rm max}\rangle$ with maximal probability.
Contrary to the case of the XXZ chain, this configuration  is non-degenerate: $d=1$ in the notation of Ref.~\cite{smp11}.
Since the Luttinger parameter $R$ is equal to $1$ for the dimer problem (free fermions), the analysis of  Ref.~\cite{smp11} immediately gives $n_c=d^2/R=1$, in agreement with the present numerics.
Above $n_c$ the universal contribution to the entropy is that of a single ``flat'' height configuration. As in the XXZ chain, this flat configuration does not correspond to a simple Dirichlet boundary condition around $A$ in the continuum limit. Indeed, the (coarse grained) height is shifted by an amount $\delta=\frac{1}{2}\pi r$ with respect to the vertical boundaries of the lattice (see Fig.~\ref{fig:shift}). As in the XXZ chain situation, this height shift produces a logarithmic term which exactly compensate the logarithmic terms coming from the Cardy-Peschel angles, hence the absence of logarithm in the R\'enyi entropy when $n\geq n_c=1$.

\section{Kitaev-Preskill construction}
\label{sec:kp}

\begin{figure}[!ht]
\begin{center}
\includegraphics{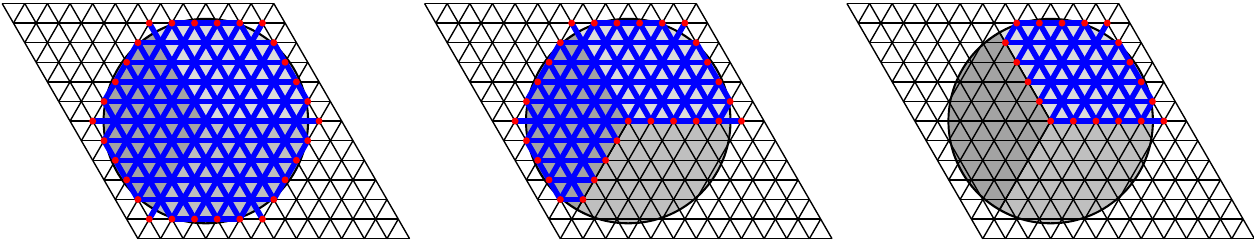}
\end{center}
\caption{Geometries required for the computation of $S_n^{\rm (ABC)}$, $S_n^{\rm (AB)}$ and $S_n^{\rm (A)}$ at Radius $\rho=4.5$. They have
 $N_b=30$, $29$ and $19$ boundary sites (in red) respectively.}
\label{fig:kp_geometry}
\end{figure}

\begin{figure}[!ht]
\includegraphics[width=7.7cm]{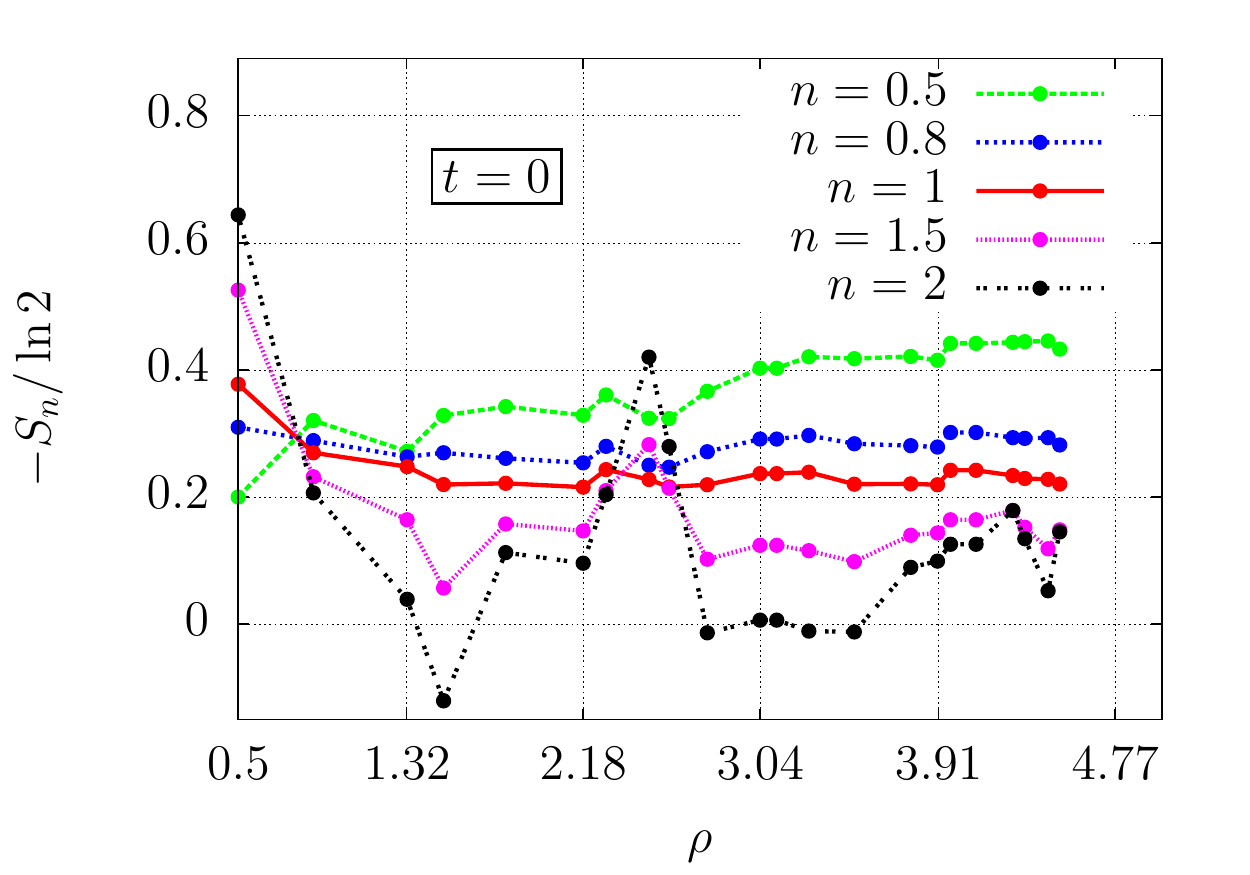}\hfill
\includegraphics[width=7.7cm]{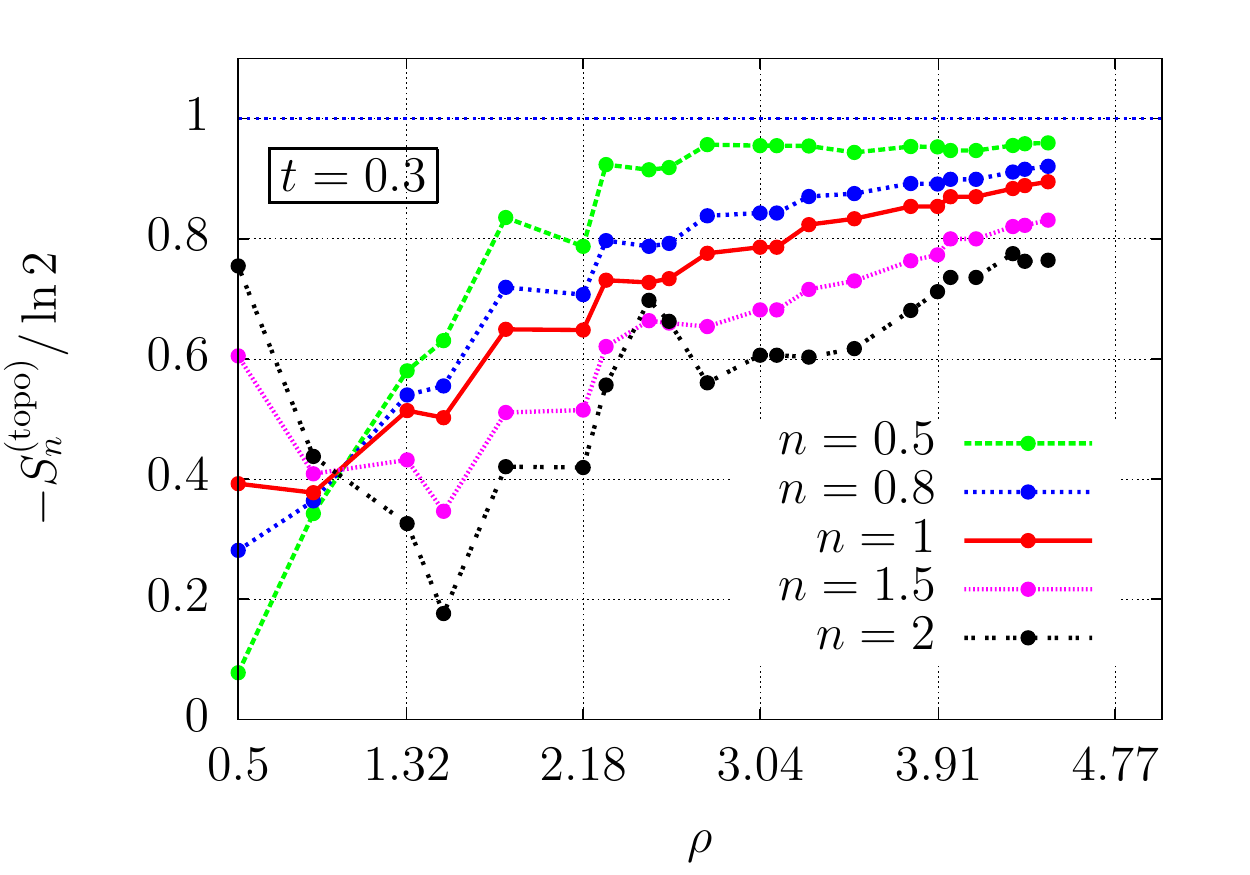}
\includegraphics[width=7.7cm]{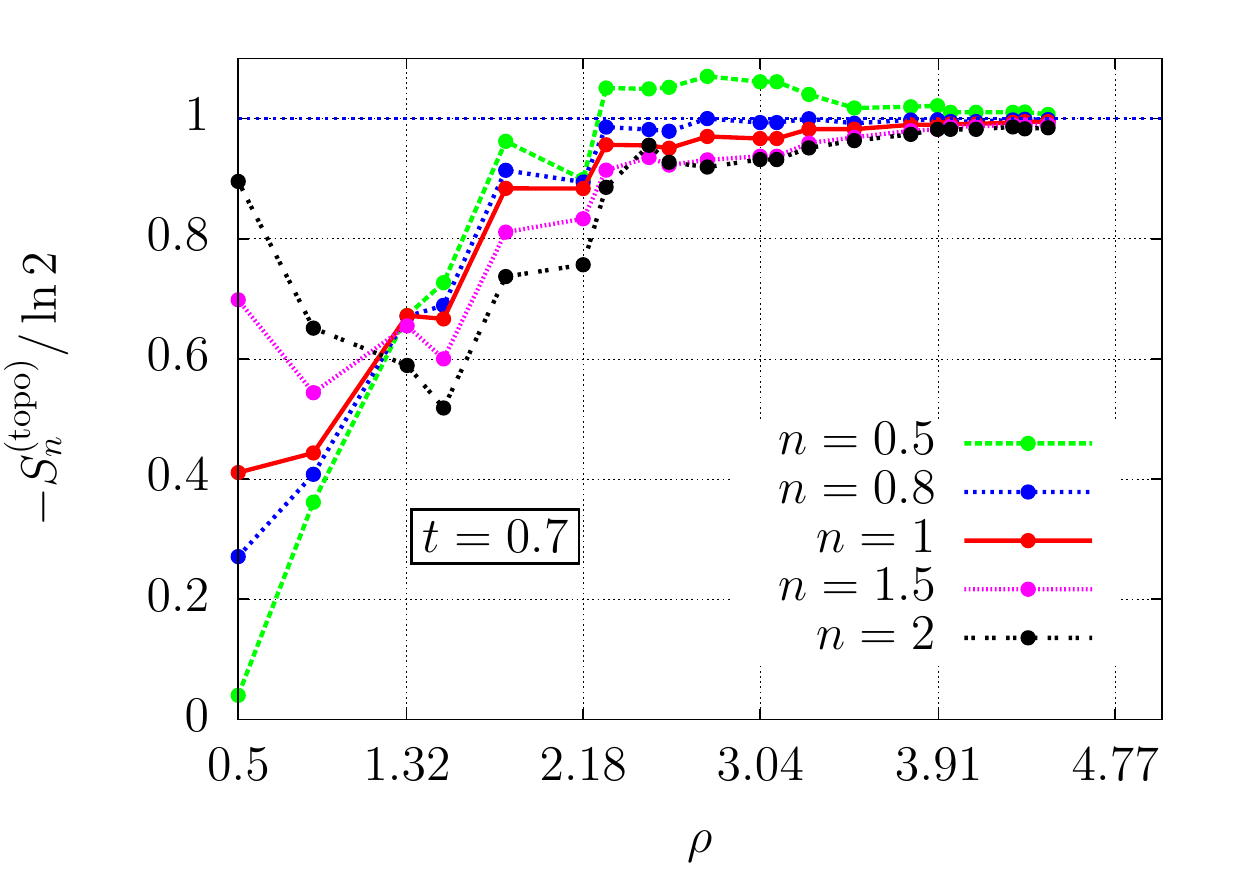}\hfill
\includegraphics[width=7.7cm]{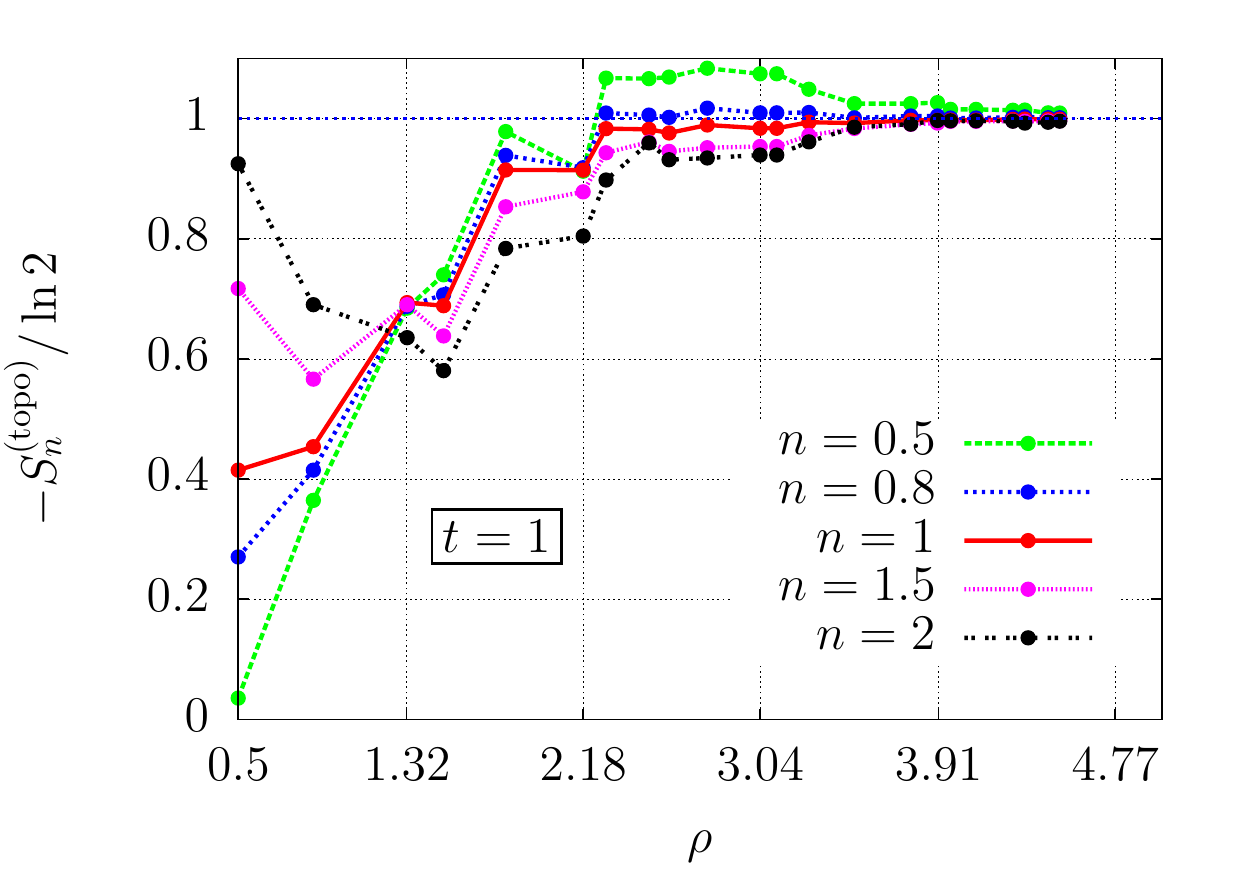} 
\caption{Top left : critical case $t=0$. Top right : $t=0.3$. Bottom left $t=0.7$. Bottom right $t=1$.
 In each case, $-S_n^{\rm topo}/\ln 2$ is shown, for $n=0.5,0.8,1,1.5,2$ as a function of the radius $\rho$.}
\label{fig:KPresults}
\end{figure}

As discussed in Sec.~\ref{sec:cyl} the cylinder geometry allows to extract the subleading entropy term in a rather straightforward way, by a simple fit of $S_n(L)$ on (at least) two system sizes. However, 
the original proposals \cite{lw06,kp06} were to extract the topological entanglement entropy from a single and large planar system. There, the subsystems
on which the entanglement entropy are computed cannot have a straight boundary and necessarily have corners, etc. These corners (as well as the curvature) also contribute to the entanglement entropy by a (non-universal) amount of order one and therefore need to be subtracted. The subtraction scheme proposed by Kitaev and Preskill  \cite{kp06} is based on the following combination of entropies (see Fig.~\ref{fig:kp_geometry}).
\begin{eqnarray}
 S_{n}^{\rm topo}&=&S_n^{(ABC)}-S_n^{(AB)}-S_n^{(BC)}-S_n^{(AC)}+S_n^{(A)}+S_n^{(B)}+S_n^{(C)} \label{eq:KPABC}
\end{eqnarray}
The first numerical implementation of this subtraction ideas was done in a the  RK dimer wave function at $t=1$ and $n=1$ \cite{fm06}. Some other recent works investigated the  $n=2$ case
using quantum Monte Carlo on a Bose-Hubbard model\cite{ihm11} and variational quantum Monte Carlo on projected spin liquid wave-functions \cite{zgv11}. Here we extend the results of Ref.~\cite{fm06} on dimer RK wave functions for several values of $t$, $n$, and with with finite areas $A$, $B$ and $C$ embedded in a {\it infinite} plane. The results are shown in Fig.~\ref{fig:KPresults}. Provided $t$ is not too small ({\it i.e.} the dimer-dimer correlation length is not too large), the Kitaev-Preskill construction gives an entropy constant equal to $-\ln(2)$ with high precision, as expected. Still, for the same numerical effort (boundary length), the convergence turns out to be slower than with the cylinder geometry (see Fig.~\ref{fig:convergence}). 

The Eq.~\ref{eq:KPABC} was originally designed to probe  {\it massive} wave-function, 
but it is also natural to consider the limit $t\to0$ where the wave function becomes critical (and restricting to $n<n_c$ for simplicity).

Each term in  Eq.~\ref{eq:KPABC} corresponds to a subsystem $\Omega=ABC,AB,\cdots$ which is topologically equivalent to a disk, but possibly with some sharp corners.
For each such subsystem, we wish to use
a formula derived in Ref.~\cite{smp11}:
\begin{eqnarray}
 S_n(\Omega)&=&\frac{1}{1-n}\left[
\ln\left(\frac{\mathcal Z_{n\kappa}}{\mathcal Z^D_{n\kappa}}\right) -n \ln\left(\frac{\mathcal Z_\kappa}{Z^D_\kappa}\right)
\right],
\label{eq:RnZZD}
\end{eqnarray}
where $\mathcal Z$ is a free-field partition function on the whole system, and $\mathcal Z^D$ is the partition function with Dirichlet boundary condition imposed at the boundary of $\Omega$ (thus disconnecting $\Omega$ and $\bar\Omega$). $\kappa$ is the bare stiffness and the first term should be evaluated with a modified stiffness $\kappa'=n\kappa$.\footnote{This formula was originally derived in the case in the case where $\Omega$ is a half infinite cylinder, but the argument in fact applies to the present geometries as well.}

By construction, the non-universal contributions proportional to the boundary length will drop out of the KP combination.
Next, we  consider the logarithmically divergent terms which come from the sharp corner contributions to the free energies.
Each corner with interior angle $\alpha$ gives a contribution $F(\alpha)=\frac{1}{24}(\frac{\alpha}{\pi}-\frac{\pi}{\alpha}) \ln (L/l_0)$ to the free energy, where $L$ is the typical scale
of the boundary and $l_0$ some microscopic cut off \cite{cp88}.  To apply the Eq.~\ref{eq:RnZZD} what needs to be computed is the free energy difference between that of the whole system, and that
where $\Omega$ and $\bar \Omega$ have been disconnected (Dirichlet boundary condition). So, in the disconnected term, a sharp corner of angle $\alpha$ in $\Omega$ will also contribute as a sharp corner of angle $2\pi-\alpha$
(in $\bar \Omega$). The contribution to $S_n$ is thus
$\delta S_n= F(\alpha)+F(2\pi-\alpha)=\frac{1}{24}(2-\frac{\pi}{\alpha}-\frac{\pi}{2\pi-\alpha})\ln((L/l_0)$, which is by construction symmetric under the exchange $\alpha \leftrightarrow 2\pi-\alpha$.
Then it is easy to check that in the spatial decomposition implied by \ref{eq:KPABC}, each angle appearing in some $+S_\Omega$ will cancel out with another one (with the same angle or its complement) in $-S_{\Omega'}$.

However, as already mentioned in Ref.~\cite{ihm11}, this is only true for the leading (logarithmically divergent) part, because there is no simple reason why the microscopic length scales $l_0$ should all be the same.
We thus expect some constant (non-divergent) and non-universal contribution to the entropy when $t=0$.

Refs.~\cite{fm06,hsu09} mentioned that the entanglement entropy of a disk $\Omega$ of radius $R$  embedded in a larger disk $\bar \Omega$
of radius $L$ could have a (very slowly) diverging term $\sim \ln(\ln(L/R))$ for a critical RK wave function.
However, in the lattice (dimer) version of the RK state we consider, it is easy to show
that the entropy must be finite when $L\to\infty$ while keeping $R$ fixed. The argument is as follows: the  (Von Neumann) entropy $S_1$
of a subsystem can be expressed using the probabilities $p_i$ of its boundary configurations:
\begin{equation}
 S_1=-\sum_{i=1}^N p_i \ln(p_i)
\end{equation}
where $N$ is the number of possible microscopic configurations at the boundary of $\Omega$.
If the boundary has a finite length $\sim R$, $N$ must be finite with $\ln N\sim R$. 
As a consequence, since the entropy is bounded by $\ln N$, 
we have $S_1 \lesssim R$. In other words, the entanglement entropy cannot exceed the boundary law for RK states. This bound does not involve the size of the outer system $\bar\Omega$,
and none of the entropies appearing in Eq.~\ref{eq:KPABC} can diverge when taking the outer system to its thermodynamic limit.
Why the argument of 
Ref.~\cite{fm06} does not apply to this quantity in lattice RK states is however unclear to us.
But in any case $S_{\rm topo}(R)$ cannot diverge when taking $L\to \infty$ at fixed $R$, whatever the lattice RK state provided
it has a finite number of states per site. This is indeed confirmed by our numerical estimations of $S_{\rm topo}(R)$
which are performed directly in the thermodynamic limit $L=\infty$ and which gives finite values for finite values $R$.
Although the system sizes ($R$) are too small to observe 
the true large-$R$ behavior for $t=0$ (square lattice), the argument above concerning the corner contribution indicate that it is very likely a non-universal number.

\section{Summary and conclusions}

Thanks to some extensive use of the Pfaffian solution of the classical (2d) dimer model, we have performed exact calculations
of the entanglement entropy and entanglement spectra of some dimer RK states on large subsystems. Using the cylinder and the Kitaev-Preskill geometries
we recovered the topological entanglement entropy of the $\mathbb{Z}_2$ phase, $-\ln(2)$, with high accuracy. As expected, this value
not only holds for the triangular lattice RK wave-function, but is in fact independent of the fugacity $t>0$. We also analyzed the scaling close to the critical point at $t=0$,
as well as the behavior for large values
of the R\'enyi index $n$. In particular, we proved for $n \to\infty$ that the sub-leading entropy constant is $-\ln(2)$.
Thanks to its translation-invariant boundary, the cylinder geometry gives smaller finite-size effects and therefore a much more precise estimation of the topological entanglement entropy
than the KP setup (for a given length of the subsystem boundary).
For this reason, it may be preferred in future numerical studies (exact diagonalization or quantum Monte Carlo)
looking for topological ground states  in realistic lattice models.

The entanglement spectra were also computed in the cylinder geometry, and the presence of a unique ground-state and a finite gap (whatever the fugacity) showed
that for these states, contrary to naive expectations,  the topological (or critical) nature of the phase is not apparent in the low-energy part of the entanglement spectrum.
Simpler $\mathbb{Z}_2$ wave functions such as that of the Toric Code\cite{kitaev03} (or that of Ref.~\cite{msp02})
do not allow to learn much about the structure of the entanglement spectrum. Indeed, in those states with vanishing correlation length all the non-zero eigenvalues of the reduced density matrix are exactly degenerate
(no $n$ dependence of the R\'enyi entropy).
From this point of view, the dimer states we consider offer an interesting compromise between  the possibility to do exact calculations on large systems and a non-trivial entanglement spectrum.
Extending these calculations to other states with richer topological structure, like string-nets wave functions \cite{lw05}, could be a promising direction of research.

\newpage
\begin{appendix}
\section[\qquad \qquad \;\; Green function elements for an infinite cylinder]{Green function elements for an infinite cylinder}
\label{sec:green}
\subsection[\qquad \qquad \; Diagonalization of the Kasteleyn matrix]{Diagonalization of the Kasteleyn matrix}
We wish to diagonalize the Kasteleyn matrix by Fourier transform for $L_y \to \infty$. To
 do so we must distinguish between two sublattices (see Fig.~\ref{fig:cyl_orientation}):
\begin{eqnarray}
 \mathcal{L}_0&=& \big\{(2x \mathbf{u_x}+y \mathbf{u_y})\;|\;0\leq x< L_x/2,0 \leq y<L_y\big\}\\
\mathcal{L}_1&=&\big\{(2x+1)\mathbf{u_x}+y\mathbf{u_y}\;|\;0\leq x< L_x/2,0 \leq y<L_y\big\}
\end{eqnarray}
We denote by $N=L_x L_y$ the number of sites. Then we define a new basis
\begin{eqnarray}
 |\mathbf{k},0\rangle&=&\frac{1}{\sqrt{N/2}}\;\displaystyle{\sum_{\mathbf{r}_0 \in \mathcal{L}_0} e^{-i \mathbf{k.r}_0}|\mathbf{r}_0\rangle}\\
|\mathbf{k},1\rangle&=&\frac{1}{\sqrt{N/2}}\;\sum_{\mathbf{r}_1 \in \mathcal{L}_1} e^{-i \mathbf{k.r}_1}|\mathbf{r}_1\rangle
\end{eqnarray}
The Kasteleyn matrix satisfies antiperiodic boundary conditions in the $x-$ direction, and since $L_y\to \infty$,
 we can also assume antiperiodic boundary conditions in the $y-$ direction. The appropriate wave-vectors are the $\mathbf{k}=k_x \mathbf{u_x}+k_y \mathbf{u_y}$
 with
\begin{eqnarray}
k_x&\in& K_x=\left\{\frac{(2j+1)\pi}{L_x}\quad \Big|\quad j=0,\ldots ,L_x/2-1\right\}\\
k_y &\in& K_y=\left\{\frac{(2j+1)\pi}{L_y}\quad \Big |\quad j=0,\ldots ,L_y-1\right\}
\end{eqnarray}
In the new basis, the Kasteleyn matrix takes the following simple form
\begin{equation}\label{eq:block}
 \mathcal{K}_{\alpha\beta}(\mathbf{k})=\left(\begin{array}{cc}2i\sin k_y& 2i\sin k_x +2 t\cos(k_x+k_y)\\ 2i \sin k_x -2 t \cos(k_x+k_y)& -2i \sin k_y  \end{array}\right),
\end{equation}
and can easily be inverted
\begin{equation}\fl
 \mathcal{K}^{-1}_{\alpha\beta}(\mathbf{k})=\frac{1}{\det \left[\mathcal{K}_{\alpha \beta}(\mathbf{k})\right]}
\left(\begin{array}{cc}-2i\sin k_y& -2i\sin k_x -2 t\cos(k_x+k_y)\\ -2i \sin k_x +2 t \cos(k_x+k_y)& 2i \sin k_y  \end{array}\right)
\end{equation}
with
\begin{equation}
 \det \left[ K_{\alpha \beta}(\mathbf{k})\right]=4\sin^2 k_x +4\sin^2 k_y +4t^2 \cos^2(k_x+k_y).
\end{equation}
For two sites $\mathbf{r}=x\mathbf{u_x}+y \mathbf{u_y}$ and $\mathbf{r'}=x'\mathbf{u_x}+y' \mathbf{u_y}$ respectively in sublattices $\alpha$ and $\beta$,
 the Green function element is
\begin{equation}\label{eq:green_elts}
 \mathcal{K}^{-1}_{\mathbf{r},\mathbf{r'}}=
\frac{1}{\pi L_x}\sum_{k_x}e^{-ik_x (x'-x)}\int_{0}^{2\pi}dk_y \, \mathcal{K}^{-1}_{\alpha \beta}(\mathbf{k})e^{-ik_y(y'-y)}
\end{equation}
In this equation, the integral on $dk_y$ can in principle be done explicitly for any $y'-y$, as will be shown in the next subsection.
 To compute the entanglement entropy in the cylinder geometry $|y'-y|$ doesn't however need to be greater than $2$,
 whereas it can attain $3$ in the strip geometry. 
\subsection[\qquad \qquad \;Green function elements]{Green function elements}
The computation of Green functions element requires the evaluation of integrals of the form
\begin{eqnarray}
 C_p(k_x)&=&\int_{0}^{2\pi} \frac{\cos (p \,k_y)}{4\sin^2 k_x +4\sin^2 k_y +4t^2 \cos^2(k_x+k_y)}d k_y\\
S_p(k_x)&=&\int _{0}^{2\pi} \frac{\sin (p \,k_y)}{4\sin^2 k_x +4\sin^2 k_y +4t^2 \cos^2(k_x+k_y)}d k_y,
\end{eqnarray}
with $p$ an even integer (otherwise the integrals are simply zero by symmetry). Both integrands are $\pi-$periodic and following
 Bioche's rules we can make the change in variables $u=\tan k_y$. We get
\begin{eqnarray}\fl
 C_p(k_x)&=&\frac{1}{2}\int_{-\infty}^{+\infty}\frac{T_p\left[(1+u^2)^{-1/2}\right]\,du}
{u^2[1+(1+t^2)\sin^2 k_x]-u t^2 \sin (2k_x)+\sin^2 k_x +t^2 \cos^2 k_x}\label{eq:int1}\\\fl
S_p(k_x)&=&\frac{1}{2}\int_{-\infty}^{+\infty}\frac{u (1+u^2)^{-1/2}\,U_{p-1}\left[(1+u^2)^{-1/2}\right]\,du}
{u^2[1+(1+t^2)\sin^2 k_x]-u t^2 \sin (2k_x) +\sin^2 k_x +t^2 \cos^2 k_x}\label{eq:int2}
\end{eqnarray}
where $T_p(x)$ and $U_{p-1}(x)$ are the Chebyshev polynomials of the first and second kind respectively:
\begin{eqnarray}
 T_p(\cos \theta)&=& \cos p\theta\\
U_{p-1}(\cos \theta)&=&\frac{\sin p\theta}{\sin \theta}
\end{eqnarray}
For $p$ even $T_p(-x)=T_p(x)$ and  $U_{p-1}(-x)=-U_{p-1}(x)$. Therefore, both integrands in Eq. \ref{eq:int1} and \ref{eq:int2}
 are {\it rational} functions of $u$, as should be. $C_p$ and $S_p$ can then be calculated by residue. Closing the contour by a big circle in the upper-half plane, 
two poles will contribute to the integral. The first pole is at 
\begin{equation}
 u=\frac{t^2 \sin k_x \cos k_x +i \sqrt{t^2+\sin^2 k_x +\sin^4 k_x}}{1+(1+t^2)\sin^2 k_x}
\end{equation}
and is of order $1$. The second one at $u=i$ is there if $p\neq 0$ and is of order $p/2$. Although the residue calculation
 for any even $p$ is in principle straightforward, the procedure becomes more and more cumbersome when $p$ gets bigger. Only for $p=0$ do we get a simple (known\cite{fms02}) result:
\begin{equation}\label{eq:simple_green}
 C_0(k_x)=\frac{\pi/2}{\sqrt{t^2+\sin^2 k_x+\sin^4 k_x}}
\end{equation}
From these we can get access to all the Green functions elements. The simplest are along the same horizontal line, and only require the knowledge of $C_0$ :
\begin{eqnarray}
\mathcal{K}^{-1}_{\,2\ell \mathbf{u}_x}&=&0\\\label{eq:greenhoriz}
\mathcal{K}^{-1}_{(2\ell+1)\mathbf{u}_x}&=&\frac{1}{L_x}\sum_{k_x} \frac{\sin k_x\sin (2\ell+1) k_x}{\sqrt{t^2+\sin^2 k_x+\sin^4 k_x}}
\end{eqnarray}
For the cylinder geometry, the knowledge of $C_0$, $C_2$ and $S_2$ is sufficient. For the strip geometry, also
$C_4$ and $S_4$ are needed. To compute the entanglement entropy in the Kitaev-Preskill geometry, it is easier to evaluate the double integral ($L_x \to \infty$)
in Eq.~\ref{eq:green_elts} numerically.

\section[\qquad \qquad \;\;Closed-form formula for $S_{n=\infty}$ in the cylinder geometry]{Closed-form formula for $S_{n=\infty}$ in the cylinder geometry}
\label{sec:renyi_infinity}
As explained in the text, the maximum probability corresponds to a simple configuration with all boundary spins up.
 Then, a natural way to proceed would be to use Eq.~\ref{eq:perturbation_trick} and try to evaluate the resulting determinant.
 This method is most certainly viable, but we will follow another path. In the dimer language, the probability we are looking for is given by
\begin{equation}\label{eq:pmax_ratio}
p_{\rm max}= \lim_{L_y \to \infty}\frac{\left[Z_{\rm cyl}(L_x,L_y/2)\right]^2}{Z_{\rm cyl}(L_x,L_y)},
\end{equation}
 where $Z_{\rm cyl}(L_x,h)$ counts the number of dimer coverings on a finite cylinder of circumference $L_x$ and height $h$. Despite the loss
 of translational invariance in the $y-$ direction, $Z_{\rm cyl}$ can still be evaluated in closed form, as is shown in \ref{sec:finite_cylinder}. From this
 $p_{\rm max}$ can easily be calculated, see \ref{sec:pmax}
\subsection[\qquad \qquad \;Dimer coverings on a finite cylinder]{Dimer coverings on a finite cylinder}
\label{sec:finite_cylinder}
Let $Z_{\rm cyl}$ be the partition we are looking for. Using (skew) translational invariance along the x-axis, one gets (recall $K_x=\{(2m-1)\pi/L_x \quad,1\leq m \leq L_x/2\}$):
\begin{equation}
 Z_{\rm cyl}(L_x,L_y)^2=\prod_{k_x \in K_x} \det \left[\mathcal{K}^{(x)}_{1\leq i,j \leq 2L_y}\right] 
\end{equation}
In other word, the Kasteleyn matrix is block-diagonal with $L_x/2$ blocks of size $2L_y$. Setting $t_x=t e^{ix}$ and $s_x=2i \sin x$, 
\begin{equation}\fl
\mathcal{K}^{(x)}=\left(\begin{array}{cccccccccccc}0&s_x&1&t_x&0&&&&&&&\\
 s_x &0 &-t_x&-1&0&&&&&&&\\
-1&\bar{t}_x&0& s_x&1&t_x&&&&&&\\
-\bar{t}_x&1&s_x & 0 & -t_x&-1&&&&&&\\
0&0&-1&\bar{t}_x&0& s_x&1&t_x&&&&\\
0&0&-\bar{t}_x&1&s_x & 0 & -t_x&-1&&&&\\
&&&&&&&&&&&\\
&&&&&&&&&&&\\
&&&&&&-1&\bar{t}_x&0& s_x&1&t_x\\
&&&&&&-\bar{t}_x&1&s_x & 0 & -t_x&-1\\
&&&&&&&&-1&\bar{t}_x&0& s_x\\
&&&&&&&&-\bar{t}_x&1&s_x & 0
\end{array}\right)
\end{equation}
Although it is not easy to diagonalize $\mathcal{K}^{(x)}$ , its determinant can be exactly evaluated using the perturbation trick. 
To do so, we introduce
\begin{equation}\fl
\mathcal{K}^{(x)}_0=\left(\begin{array}{cccccccccccc}0&s_x&1&t_x&0&&&&&&-1&-\bar{t}_x\\
 s_x &0 &-t_x&-1&0&&&&&&\bar{t}_x&1\\
-1&\bar{t}_x&0& s_x&1&t_x&&&&&&\\
-\bar{t}_x&1&s_x & 0 & -t_x&-1&&&&&&\\
0&0&-1&\bar{t}_x&0& s_x&1&t_x&&&&\\
0&0&-\bar{t}_x&1&s_x & 0 & -t_x&-1&&&&\\
&&&&&&&&&&&\\
&&&&&&&&&&&\\
&&&&&&-1&\bar{t}_x&0& s_x&1&t_x\\
&&&&&&-\bar{t}_x&1&s_x & 0 & -t_x&-1\\
-1&-t_x&&&&&&&-1&\bar{t}_x&0& s_x\\
t_x&1&&&&&&&-\bar{t}_x&1&s_x & 0
\end{array}\right)
\end{equation}
This amounts to putting antiperiodic boundary condition along the $y-$ axis for the total Kasteleyn matrix. $\mathcal{K}_0^{(x)}$ is block skew circulant, and
 it can be diagonalized in Fourier space. In particular its determinant can be easily evaluated : 
\begin{eqnarray}
 \det \mathcal{K}_0^{(x)}&=&\prod_{k_y\in K_y}\Delta(k_x,k_y)\\
 \Delta(k_x,k_y)&=&4\sin^2 k_x +4\sin^2 k_y +4t^2 \cos^2 (k_x+k_y),
\end{eqnarray}
where $K_y=\{(2m-1)\pi/L_y\quad,1\leq m \leq L_y\}$. This allows to express $\det \mathcal{K}^{(x)}$ as
\begin{equation}
 \frac{\det \mathcal{K}^{(x)}}{\det \mathcal{K}_0^{(x)}}=\det \left(1+\left[\mathcal{K}_0^{(x)}\right]^{-1} \left[\mathcal{K}^{(x)}-\mathcal{K}_0^{(x)}\right]\right)
=\det M_4^{(x)}
\end{equation}
$\mathcal{K}^{(x)}-\mathcal{K}_0^{(x)}$ is a matrix with only 8 non-zero elements, and using elementary row-column manipulations, the determinant can be reduced
 to a $4\times 4$:
\begin{equation}
 M_4^{(x)}=\left(
\begin{array}{ccccc}
 z&-a&w&-ib\\
-a&z&ib&-w\\
-\bar{w}&ib&\bar{z}&a\\
-ib&\bar{w}&a&\bar{z}
\end{array}
\right)
\qquad (z,w,a,b)\in \mathbb{C}\times\mathbb{C}\times \mathbb{R}\times \mathbb{R}
\end{equation}
After some algebra, we get the following formulae for the coefficients :
\begin{eqnarray}
z&=&\frac{1}{2}+\frac{2}{L_y}\sum_{k_y} \frac{\sin^2 k_x+i\left[\sin(2k_y)-t^2\sin(2k_x+2k_y)\right]}{\Delta(k_x,k_y)}\\
a&=&\frac{2t}{L_y}\sum_{k_y} \frac{\cos k_x}{\Delta(k_x,k_y)}\\
w&=&\frac{2it}{L_y}\sum_{k_y} \frac{\sin k_x \,e^{-ik_x}}{\Delta(k_x,k_y)}\\
b&=&\frac{2}{L_y}\sum_{k_y} \frac{\sin k_x}{\Delta(k_x,k_y)}
\end{eqnarray}
The number of dimer coverings on the triangular lattice with cylindrical boundary conditions is then given by:
\begin{equation}\label{eq:cyl_dimers}
 Z_{\rm cyl}(L_x,L_y)=\prod_{k_x}\left\{ \det\left(M_4^{(x)}\right)\times \prod_{k_y} \Delta(k_x,k_y)\right\}^{1/2}
\end{equation}
Evaluating the determinant, we finally get the following closed formula for the partition function
\begin{equation}
 Z_{\rm cyl}(L_x,L_y)=\prod_{k_x}\left\{
A(k_x) \times\prod_{k_y}
\left[
\Delta(k_x,k_y)
\right]^{1/2}
\right\},
\end{equation}
with
\begin{eqnarray}\fl\nonumber
 A(k_x)&=&\left(t^2+\sin^2 k_x+\sin^4 k_x\right)d(k_x)^2+(\sin^2 k_x-t\cos k_x) d(k_x) +1/4+\varepsilon(k_x)^2\\\fl\nonumber
d(k_x)&=&\frac{2}{L_y}\sum_{k_y} \frac{1}{\Delta(k_x,k_y)}\\\fl
\varepsilon(k_x)&=&\displaystyle{\frac{2}{L_y}\sum_{k_y} \frac{\sin(2k_y)-t^2 \sin(2k_x+2k_y)}{\Delta(k_x,k_y)}}.
\end{eqnarray}
\subsection[\qquad \qquad \;Exact formula for $S_{\infty}$]{Exact formula for $S_{n=\infty}$}
\label{sec:pmax}
The maximum probability is in the thermodynamic limit given by
\begin{eqnarray}\label{eq:pmax1}
 p_{\rm max}&=&\lim_{L_y \to \infty}\frac{\left[Z_{\rm cyl}(L_y/2,L_x)\right]^2}{Z_{\rm cyl}(L_y,L_x)}\\\label{eq:pmax2}
&=&\prod_{k_x}  \left(\lim _{L_y \to \infty}A(x)\right)
\end{eqnarray}
Eq.~\ref{eq:pmax2} follows from Eq.~\ref{eq:pmax1} using Euler-Maclaurin's formula on the ratio of terms involving $\Delta(k_x,k_y)$, coming from
Eq.~\ref{eq:cyl_dimers}.
Using Eq.~\ref{eq:simple_green}, we also have
\begin{equation}
\lim_{L_y \to \infty} d(k_x)=\frac{1}{2\sqrt{t^2+\sin^2 k_x+\sin^4 k_x}},
\end{equation}
while $\displaystyle{\lim_{L_y \to \infty}} \varepsilon(k_x)=0$ because the integrand has a symmetry center solution
 of $\sin (2k_y)=t^2 \sin (2k_x+2k_y)$. In the end we obtain
\begin{equation}\label{eq:logpmax}
 S_{\infty}=-\ln p_{\rm max}=-\sum_{k_x=\frac{(2m-1)\pi}{L}}^{1\leq m \leq L/2}
\ln \left(  \frac{1}{2} +\frac{1}{2}\frac{\sin^2 k_x-t \cos k_x}{\sqrt{t^2+\sin^2 k_x+\sin^4 k_x}}\right)
\end{equation}
\subsection[\qquad \qquad \;Asymptotic expansion]{Asymptotic expansion}
At $t=0$, the subleading constant in the $L\to\infty$ asymptotic expansion just follows from the Euler-Maclaurin formula. We find
\begin{equation}
 s_\infty(t=0)=0.
\end{equation}
Some additional care must be taken in the case $t>0$. The function
\begin{equation}
 f(k)=-\ln \left(  \frac{1}{2} +\frac{1}{2}\frac{\sin^2 k-t \cos k}{\sqrt{t^2+\sin^2 k+\sin^4 k}}\right)
\end{equation}
actually diverges as $f(k)\sim-2\ln k$ -- independent on $t$ -- when $k\to 0$. The asymptotics can be obtained by applying the Euler-Maclaurin formula on $\sum_k \left[f(k)+2\ln k\right]$ while applying Stirling's formula on the remaining ``linearized'' term $-\sum_k 2\ln k$. Doing so we finally obtain the topological term
\begin{equation} 
s_\infty(t>0)=-\ln 2.
\end{equation}
Only the linearized term actually contributes to the constant. Indeed, it is universal and shouldn't be affected by 
the short-distance ({\it i.e} high momentum $k$) details of the model.  
\end{appendix}

\section*{References}


\end{document}